\documentclass[10pt,twocolumn,journal]{IEEEtran}
\usepackage{subfigure}
\usepackage{graphicx}
\usepackage{amssymb}
\ifCLASSINFOpdf
\else
\fi
\begin{document}
%
\title{Spatial multi-LRU: Distributed Caching for Wireless Networks with Coverage Overlaps}
%
%
%

\author{Anastasios~Giovanidis, 
        Apostolos~Avranas
\thanks{A. Giovanidis (anastasios.giovanidis@lip6.fr) conducted this research while in the CNRS-LTCI laboratory, T\'el\'ecom ParisTech, 23 avenue d'Italie, 75013, Paris, France; he is now with the CNRS-LIP6 laboratory of the University Pierre et Marie Curie (UPMC), Sorbonne Universities, 4 place Jussieu, Boite courrier 169, Couloir 26-00, Bureau 107, 75252 Paris Cedex 05, France.
}
\thanks{A. Avranas (apostolos.avranas@huawei.com) conducted part of this research as student of Aristotle University of Thessaloniki, Greece, and of T\'el\'ecom ParisTech, Paris, France. He is currently with the Mathematical and Algorithmic Sciences Lab, France Research Center, Huawei Technologies Co. Ltd., 
Arcs de Seine Bat. A, 20 quai du Point du Jour 92100 Boulogne Billancourt, France.
}
\thanks{Preliminary versions of this material have been presented at ACM SIGMETRICS '16 / IFIP Performance \cite{GiovaSIGM16} and IEEE ISTC '16 \cite{AvraISTC16}.}
}

\maketitle


%
\IEEEpeerreviewmaketitle

\begin{abstract}
This article introduces a novel family of decentralised caching policies, applicable to wireless networks with finite storage at the edge-nodes (stations). These policies, that are based on the \textit{Least-Recently-Used} replacement principle, are here referred to as spatial \textit{multi-LRU}. They update cache inventories in a way that provides content diversity to users who are covered by, and thus have access to, more than one station. Two variations are proposed, the \textit{multi-LRU-One} and \textit{-All}, which differ in the number of replicas inserted in the involved caches. We analyse their performance under two types of traffic demand, the Independent Reference Model (IRM) and a model that exhibits temporal locality. For IRM, we propose a Che-like approximation to predict the hit probability, which gives very accurate results. Numerical evaluations show that the performance of multi-LRU increases the more the multi-coverage areas increase, and it is close to the performance of centralised policies, when multi-coverage is sufficient. For IRM traffic, multi-LRU-One is preferable to multi-LRU-All, whereas when the traffic exhibits temporal locality the -All variation can perform better. Both variations outperform the simple LRU. When popularity knowledge is not accurate, the new policies can perform better than centralised ones.
\end{abstract}

\begin{IEEEkeywords}
Wireless, Cache, LRU, Information Centric Networking, Hit Probability, Popularity, IRM, Temporal locality.
\end{IEEEkeywords}




\section{Introduction}

\IEEEPARstart{T}{he} design of today's and future networks is characterised by a paradigm shift, from a host-centric communication
architecture, towards an Information Centric Networking (ICN) one \cite{ICNkatsaros14}. 
Following this novel concept, network nodes are equipped with storage capacity where data objects can be temporarily cached and retrieved \cite{Paschos1602}. In this way information can be made available closer to the user, it can be accessed reliably \cite{Timo15} with minimum delay, and possibly with a quality adaptable to the users' preferences, as envisioned in the case
of multimedia files. The principal benefits are the partial elimination of redundant traffic 
flows at the core network by serving demands from intermediate nodes, as well as reduced latency of service \cite{Sourl11}. 
The edge-nodes constitute a very important part of the ICN architecture, since it is where the wireless users directly have access
to. When these nodes are equipped with storage capability, download path length is minimised \cite{ElRobSIGCOMM15}. 

In this work, we consider the wireless edge of a content centric network, which consists of a set of transmitting nodes taking
fixed positions on a planar area, and a set of users dynamically arriving at this area and asking for service. The set
of transmitters can refer to base stations (BSs) of a cellular network, small stations of heterogeneous networks, WIFI
hotspots, or any other type of wireless nodes that can provide access to users. 
A user can be covered by multiple of these nodes, but she/he will be served by only one. Each node is equipped with memory of some given size.

An important question is how to best manage the available edge-memories, in order to maximise the hit probability of user-demands. We define the hit probability as the probability that a user finds her/his demand cached in
the memory of one of the covering cells. By managing, we mean to choose a policy that decides which
objects to install in each cache and how each cache inventory is updated over time. 

Given the possibility for multi-coverage, cache management should target two, somewhat conflicting, goals:
On the one hand make popular objects, which attract the large bulk of demands, generously available at many geographic
locations. On the other hand, make good use of multi-coverage, by filling the memory caches in a way that provides large object variety to each user, so that also less popular objects can be found in the caches.
Additionally, since wireless nodes (BSs) are scattered over a very large area and are of considerable number, related operations should be distributed as in \cite{MassoulieCache15, BorstINFOCOM10, LeconteITC15}, and centralised solutions should be avoided.


\subsection{Related research} 

\textit{Single Cache:} There exists a variety of cache placement policies that apply to \textit{single caches}, when no coverage overlap is considered. These include the Least Frequently Used (LFU), the Least Recently Used (LRU), and their variations. Specifically LRU has been extensively studied and approximations to the hit probability have been proposed, like the one from Dan and Towsley \cite{DanTow90}. Che et al proposed in 2002 \cite{CheApprox02} a simple approximation for the (single-)LRU under the Independent Reference Model (IRM) \cite{CoffmanBook73}, which results in an analytic formula for the hit probability with excellent fit to simulations. This fitness is theoretically explained by Fricker et al in \cite{FriRoRo12}. Application of the Che approximation under more general traffic conditions, to variations of the LRU for single caches as well as networks of caches, is provided by Martina et al \cite{LeoINFOCOM14}. In that work, and further in Elayoubi and Roberts \cite{ElRobSIGCOMM15}, it is shown that for mobile networks, application of pre-filtering improves the performance of LRU. 

\textit{Multiple Caches:} The problem of optimal content placement, when network areas are covered by more than one station has also been recently studied. A number of pro-active caching policies have been proposed, where the cache inventories are pre-filled by content, based on knowledge of the content popularity distribution and additional network-related information. Golrezaei et al \cite{GolrezaeiINFOCOM12} find the optimal content placement that maximises hit probability, when full network information (popularity, node and user positions) is available. They formulate a binary optimisation problem and propose approximation and greedy algorithms for its solution. Using reduced information (content popularity, coverage probability), B{\l}aszczyszyn and Giovanidis \cite{BlaGioICC15} provide a randomised strategy that maximises the hit probability. Poularakis et al. \cite{PoulTCOM14} formulate and solve the joint content placement and user association problem that maximises the fraction of content served by the caches of the edge-nodes. Araldo et al. \cite{AraldoSIGC14} propose joint cache sizing/object placement/path selection policies that consider also the cost of content retrieval. Recently, Naveen et al. \cite{MassoulieCache15} have formulated the problem in a way to include the bandwidth costs, and have proposed an online algorithm for its solution. Further distributed replication strategies that use different system information are proposed by Borst et al \cite{BorstINFOCOM10}, and also by Leconte et al \cite{LeconteITC15}. The problem of optimal request routing and content caching for minimum average content access delay in heterogeneous networks is studied by Dehghan et al in \cite{DehgINFO15}.

\textit{Traffic:} There can be strong dependencies between content demands, objects can have a finite lifespan, and new ones can appear anytime. These phenomena constitute the \textit{temporal locality} (not captured from the standard IRM traffic model). Such type of traffic was studied for (single-)LRU initially by Jelenkovi{\'c} and Radovanovi{\'c} \cite{JelenDep04}, and recently using also statistics from user measurements, by Traverso et al \cite{TraversoTranMult15} and Olmos et al \cite{OlmosTEMPO14}.

\textit{Point Processes:} The cache management problem for cellular networks has also been approached using point process modelling of the network node positions. Bastug et al. \cite{BastugEURASIP15} find the outage probability and content delivery rate for a given cache placement. Furthermore, Tamoor-il-Hassan et al \cite{BennisISWCS15}  find the optimal station density to achieve a given hit probability, using uniform replication. The policy in \cite{BlaGioICC15} can also  be applied for point process BS placement. 


\subsection{Contributions} 

This work has the following contributions to the subject of caching at the network edge.

- \textit{Main contribution:} It introduces (Sec. \ref{Sec:2_Cache}) a new family of decentralised caching policies, which exploit multi-coverage, called \textit{spatial multi-LRU}. Specifically, two variations of this family are studied, namely multi-LRU-One and -All. These policies constitute an extension of the classic (single-)LRU, to cases where objects can be retrieved by more than one cache. The work investigates how to best adapt the actions of update, insertion and eviction of content for multiple caches.

- The modelling takes geometry and time explicitly into consideration for the analysis of caching policies. Specifically, it investigates a three-dimensional model (two-dimensional space and time). In this, stations have a certain spatial distribution (modelled by Point Processes) and coverage areas may overlap, allowing for multi-coverage. Furthermore, it is a dynamic model, where users with demands arrive over time at different geographic locations (Sec. \ref{Sec:3_Network}).

- The hit probability performance of the new policies is evaluated for two types of traffic: (a) IRM (Sec. \ref{Sec:3_IRM}), and (b) traffic with temporal locality (Sec. \ref{Sec:4_Temp}). Specifically for the case of IRM, we initiate from the Che approximation to derive new analytic solutions (Sec. \ref{GenChe}). Two additional approximations are made here, namely the Cache Independence Approximation (CIA) for multi-LRU-One, and the Cache Similarity Approximation (CSA) for multi-LRU-All, that allow for simple but accurate analytic formulas.

- The performance of the policies is numerically evaluated. Verification for the Che-like approximations (IRM), as well as further comparison of the multi-LRU policies with other policies from the literature, under both traffic inputs, are provided in Sec. \ref{Sec:5_Simul}. For comparison we consider distributed as well as centralised policies that use various network information. 

\textit{Important conclusions:} For IRM, the multi-LRU-One always performs better than the -All variation. For traffic with temporal locality, the multi-LRU-All can perform better than -One in cases where sufficient memory is available. Both policies outperform the (single-)LRU and perform close to centralised policies for IRM traffic with significant multi-coverage. In the case of temporal locality it is shown that multi-LRU can better adapt to popularity changes compared to policies which depend on popularity estimates and content prefetching.


%
\section{Caching and its Management}
\label{Sec:2_Cache}

Caching policies can profit from various system information related e.g. to user traffic, node positions and coverage areas, or caching decisions of neighbouring nodes. Specifically, based on the available knowledge on content popularity, cache management policies can be grouped into two categories:

(i) \textit{Policies with per-reQuest updates} (POQ). For these, file popularity information is not available. Updates of the cache content are done locally and are triggered by the users on a per-request basis. The Least Frequently Used (LFU), as well as the Least Recently Used (LRU and q-LRU) policies for single cache fall in this category. 

How \textbf{LRU} works: Given an isolated cache of size $K$, the policy keeps the $K$ most recently demanded objects. The first position of the cache is called Most Recently Used (MRU) and the last one Least Recently Used (LRU). When a new demand arrives, there are two options. (a. \textit{Update}) The object demanded is already in the cache and the policy updates the object order by moving it to the MRU position. (b. \textit{Insertion}) the object is not in the cache and it is inserted as new in the MRU position, while the object in the LRU position is \textit{evicted}. In this work we call this policy, \textit{single-LRU}.

(ii) \textit{Policies with Popularity updates} (POP), where exact or estimated information over content popularities is available, and is used to infrequently update cache inventories by prefetching. This category covers the Most Popular Content (MPC) caching strategy, as well as policies that result from solutions of optimisation problems with a-priori knowledge of additional system information, e.g. the \textit{Greedy Full Information (GFI)} \cite{GolrezaeiINFOCOM12}, and the \textit{Probabilistic Block Placement (PBP)} \cite{BlaGioICC15}. Due to the extra information, POP are expected to have higher hit-probability than POQ (but this is not always true).

\subsection{Spatial multi-LRU}

This work introduces a novel family of distributed cache management POQ policies that profit from multi-coverage. These are the \textit{spatial multi-LRU} policies and are based on the single-LRU. The main idea is that, since a user can check all the caches of covering BSs for the demanded object and download it from any one that has it in its inventory, cache updates and object insertions can be done more efficiently than by just applying single-LRU independently to all caches. The fact that the user triggers a cache's update/insertion action, allows each cache to be indirectly informed about the inventory content of its neighbours. Variations of the multi-LRU family differ in the number of inserted contents in the network, after a missed content demand. Differences can also appear in the update phase. 

$\bullet$ \textbf{multi-LRU-One:} Action is taken in only \textit{one} cache out of the covering $m\geq 1$.  (a. Update) If the content is found in a non-empty subset of the $m$ caches, only one cache from the subset is used for download and, for this, the content can be moved to the MRU position. (b. Insertion) If the object is not found in \textit{any} cache, it is inserted only in one, while its least-recently-used object is evicted. This one cache can be chosen as the closest to the user, a random one, or from some other criterion. (Here, we choose the \textit{closest} node).

$\bullet$ \textbf{multi-LRU-All:} Insertion action is taken in \textit{all} $m$ caches. (a. Update) If the content is found in a non-empty subset of the $m$ caches, all caches from this subset are updated. (b. Insertion) If the object is not found in \textit{any} cache, it is inserted in all $m$. A variation based on q-LRU can be proposed, where the object is inserted in each cache with probability $q>0$.

The motivation behind the different versions of the multi-LRU policies is the following. When a user has more than one opportunity to be served due to multi-coverage, she/he can benefit from a larger cache memory (the sum of memory sizes from covering nodes.). In this setting, the optimal insertion of new content and update actions are not yet clear. If multi-LRU-One is applied, a single replica of the missed content is left down in one of the $m>1$ caches, thus favouring diversity among neighbouring caches. If multi-LRU-All is used, $m$ replicas are left down, one in each cache, thus spreading faster the new content over a larger geographic area (the union of $m$ covering cells), at the cost of diversity. q-multi-LRU-All is in-between the two, leaving down a smaller than $m$ number of replicas. A-priori, it is unclear which one will perform better with respect to hit probability. 

The performance largely depends on the type of incoming traffic. For fixed object catalogue and stationary traffic, diversity in the cache inventories can be beneficial, whereas for time-dependent traffic with varying catalogue, performance can be improved when many replicas of the same object are available, before its popularity perishes. 


\section{Network Model}
\label{Sec:3_Network}

\textit{Wireless multi-coverage}: For the analysis, the positions of transmitters coincide with the atoms from the realisation of a 2-dimensional \textit{stationary} Point Process (PP), $\Phi_b=\left\{x_i\right\}$, indexed by $i\in\mathbb{N}_+ =\left\{1,2,\ldots\right\}$, with intensity $\lambda_b>0$ in $[m^{-2}]$. In this setting, the type of PP can be general, however we consider here:

- A homogeneous \textit{Poisson PP} (PPP) $\Phi_{b,P}$ with intensity measure $\mathbb{E}\left[\Phi_{b,P}(A)\right] = \lambda_b |A|$, for some area $A\subset \mathbb{R}^2$, where $|A|$ is the surface of $A$.

- A \textit{square lattice} $\Phi_{b,L} = \eta\mathbb{Z}^2+u_L$, $\mathbb{Z}=\left\{\ldots,-1,0,1,\ldots\right\}$, whose nodes constitute a square grid with edge length $\eta>0$, randomly translated by a vector $u_L$ that is uniformly distributed in $\left[0,\eta\right]^2$ (to make $\Phi_{b,L}$ stationary). Its intensity is equal to $\lambda_b=\eta^{-2}$. 

There are two different planar areas (\textit{cells}) associated with each atom (BS) $x_i$. The first one is the \textit{Voronoi cell} $\mathcal{V}(x_i)\subset\mathbb{R}^2$. Given a PP, the Voronoi tessellation divides the plane into \textit{non-overlapping} planar subsets, each one associated with a single atom. A planar point $z$ belongs to $\mathcal{V}(x_i)$, if atom $x_i$ is the closest atom of the process to $z$. In other words, $\mathcal{V}(x_i)=\left\{z\in\mathbb{R}^2: \left|z-x_i\right|\leq \left|z-x_j\right|,\ \forall x_j\in\Phi\right\}$.

The second one is the \textit{coverage cell} $\mathcal{C}_i$. Each transmitter node $x_i\in\Phi_b$ has a possibly random area $\mathcal{C}_i$ of wireless coverage associated with it. When users arrive inside the coverage cell of $x_i$ they can be served by it, by downlink transmission. In general $\mathcal{C}_i$ is different from $\mathcal{V}(x_i)$. Coverage cells can overlap, so that a user at a random location may be covered by multiple BSs, or may not be covered at all. The total coverage area from all BSs with their coverage cells is
$\Psi = \bigcup_{i\in\mathbb{N}_+}\{x_i+\mathcal{C}_i\}$ (see \cite[Ch.3]{BacBlaVol1}).

Due to stationarity of the PP $\Phi_b$, any planar location $y\in\mathbb{R}^2$ can be chosen as reference for the performance evaluation of the wireless model. This reference is called the \textit{typical location} $o$, and for convenience we use the Cartesian origin $(0,0)$. 

Because of the random realisation of the BS positions and the random choice of the reference location $o$, the number of BS cells covering $o$ is also random. This \textit{coverage number} random variable (r.v.) $\mathcal{N}$ (see \cite{BlaGioICC15}, \cite{KeelerBartek13}) depends on the PP $\Phi_b$ and the downlink transmission scheme, with mass function 
\begin{eqnarray}
\label{pm}
p_m := \mathbb{P}\left[\mathcal{N}=m\right], & & m=0,1,\ldots,M,
\end{eqnarray}
where $M\in\mathbb{N}_+\cup\left\{\infty\right\}$. It holds, 
$\sum_{m=1}^{M}p_m = 1$.

The choice of the coverage model determines the shape of the coverage cells and consequently the values of the coverage probabilities $p_m$. In this work the choice of $\mathcal{C}_i$ is left to be general (for the evaluation, specific models are considered). Special cases include: (1) the \textit{$\mathrm{SINR}$ Model} and (2) the \textit{$\mathrm{SNR}$ or Boolean Model}. Both models consider the coverage cell $\mathcal{C}_i$ of $x_i$, as the set of planar points for which the received signal quality from $x_i$ exceeds some threshold value $T$. The motivation is that T is a predefined signal quality, above which the user gets satisfactory Quality-of-Service. The difference between these two is that the $\mathrm{SINR}$ model refers to networks with interference (e.g. when BSs serve on the same OFDMA frequency sub-slot), whereas the $\mathrm{SNR}$ model, to networks that are noise-limited (e.g. when neighbouring BSs operate on different bandwidth, by frequency reuse). For the Boolean model the $\mathcal{C}_i$ is a ball $\mathcal{B}(x_i,R_b)$ of fixed radius $R_b$ centred at $x_i$. It coincides with the $\mathrm{SNR}$ model, when no randomness of signal fading over the wireless channel is considered. 
A more detailed presentation of the different coverage models can be found in \cite{BacBlaVol1}, \cite{BartekPIMRC13} and \cite{BacBla01}.

\textit{Storage:} We consider the case where a cache memory of size $K\geq 1$ is installed and available on each transmitter node $x_i$ of $\Phi_b$. (All content files are considered of equal size, see below). The memory inventory of node $x_i$ at time $t$ is denoted by $\Xi_i(t)$ and is a (possibly varying over time) subset of the content catalogue $\mathcal{F}(t)$, with number of elements $|\Xi_i(t)|\leq K$.

\textit{Traffic Models:} Each user arrives at some point in space and time, with a request for a specific data object. The arrivals are assumed spatially independent. We model the users by a \textit{marked space-time} Point Process in $\mathbb{R}^2\times \mathbb{R}\times \mathbb{N}$, $\Phi_u=\left\{\left(\psi_i,t_i,z_i\right)\right\}$, where $\psi_i$ takes values on the Euclidean plane, and the time $t_i$ of arrival occurs at some point on the infinite time axis. The mark $z_i$ takes as values the indices of the files/objects $j:c_j\in\mathcal{F}(t)$. Service time is considered fixed and equal to unity but it will not play any role in the analysis. In this work, we evaluate the caching policies under the following two traffic models:

- A spatially homogeneous version of the \textit{Independent Reference Model} (IRM) \cite{FaPrSIAM78} (Sec. \ref{Sec:3_IRM}).

- Traffic that exhibits \textit{temporal locality}, like the Shot Noise Model (SNM) \cite{LeoINFO15} (Sec. \ref{Sec:4_Temp}).

\textit{Typical user:} The network performance is evaluated at the \textit{typical user} $u_o$, who - due to stationarity of the PPP - will be representative of any user of the process. We suppose that this user appears at the Cartesian origin $(0,0)$, at time $t_o=0$. 
In this way, the \textit{typical user} coincides with the \textit{typical location}  $o$ of the process $\Phi_b$ at time $t=0$. 

\begin{figure*}[ht!]   
\centering  
\label{VoronoiStations}
\subfigure[Poisson Transmitters/Boolean Coverage]{
	  \centering
           \includegraphics[trim=1cm 1.6cm 1cm 1cm, clip, width=0.32\textwidth]{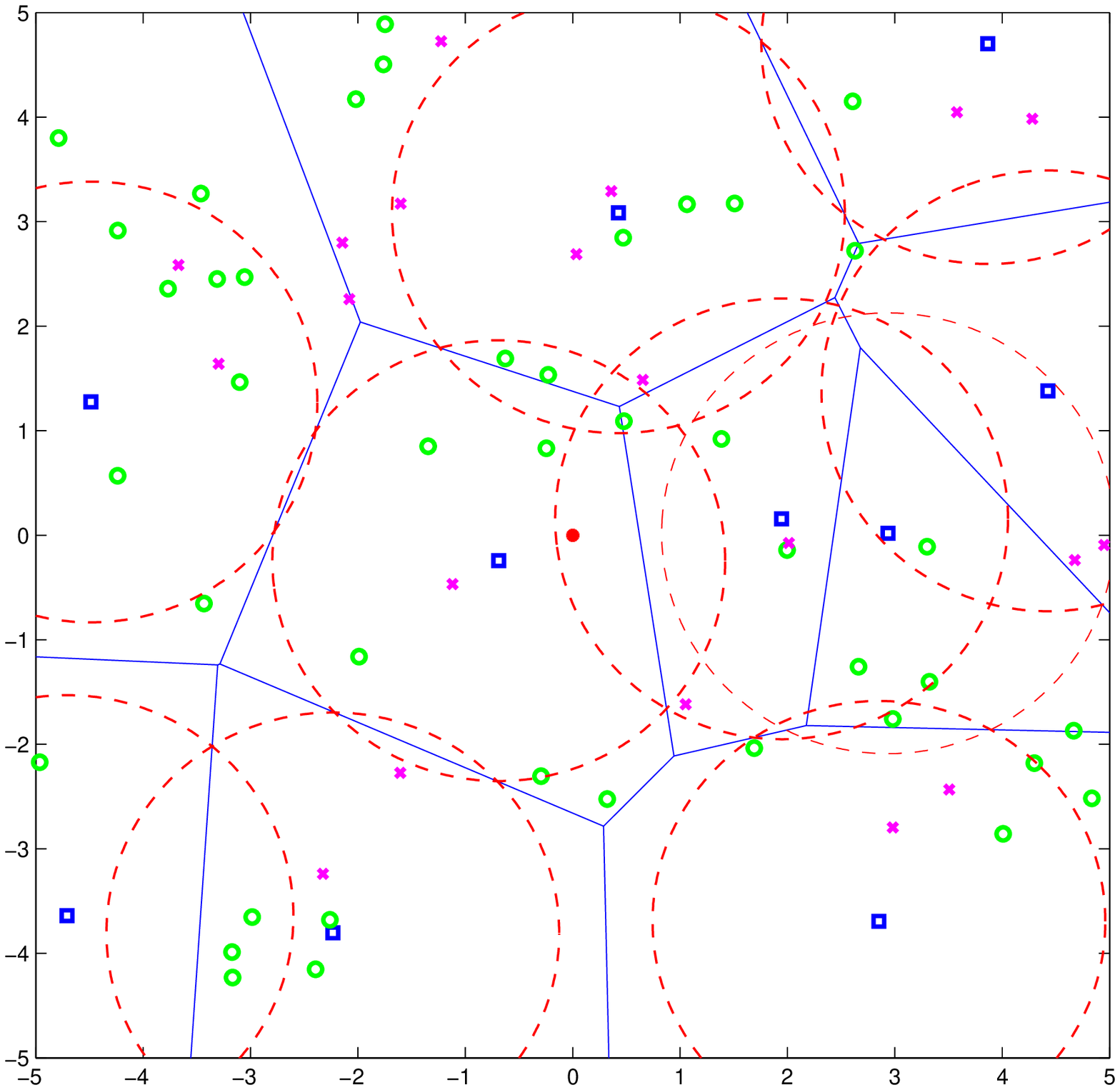}
           \label{VoronoiStations:Poisson}
           }
           \subfigure[Lattice Transmitters/Boolean Coverage]{
	   \centering  
           \includegraphics[trim=1cm 1.6cm 1cm 1cm, clip, width=0.32\textwidth]{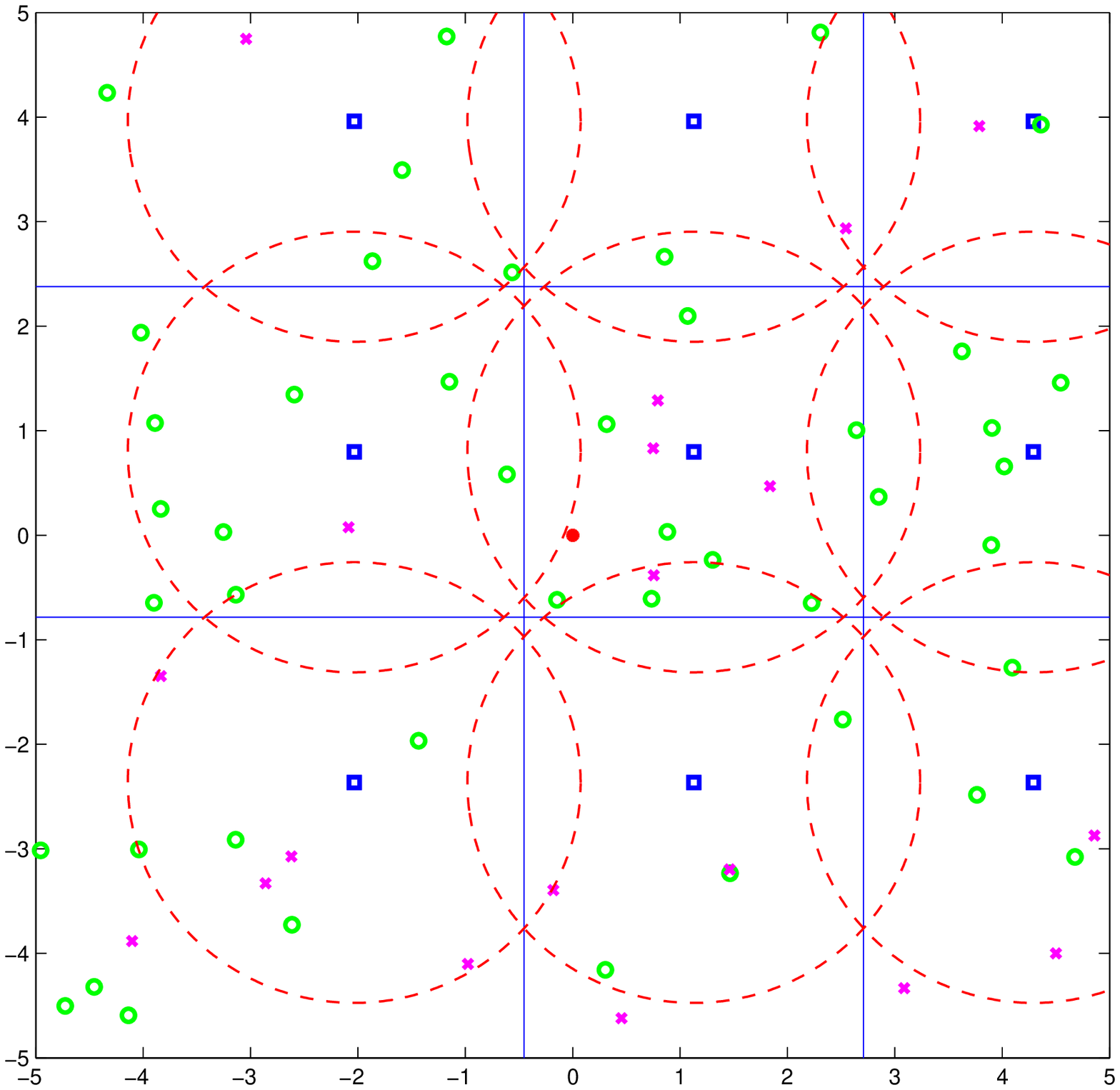}
	   \label{VoronoiStations:Lattice}
           }
\caption{A realisation of the introduced model for $t=0$ and a window of size $10\times 10$ $[m^2]$. In both subfigures, user arrivals are modelled by a PPP with $\lambda_u=0.6$ $[m^{-2} sec^{-1}]$. The users choose between two objects that have popularities $a_1=0.65$ (users with "o"), $a_2=0.35$ (users with "x"). The typical user is shown at the Cartesian origin $(0,0)$ (thicker "o"). (a) The transmitters (squares) are modelled by a PPP with $\lambda_b=0.1$ $[m^{-2}]$. (b) The transmitters (squares) are modelled by a Square Lattice PP with $\eta = \lambda_b^{-1/2}=1/\sqrt{0.1}$ $[m]$. In both figures, we assume the Boolean model for coverage, with $R_b=2\eta/3$ $[m]$. In this realisation, the typical user is covered by two cells in the PPP case and by a single one in the Lattice case. }
\end{figure*}

The model described so far is illustrated in Fig.\ref{VoronoiStations:Poisson} for the case of Poisson placement of transmitters $\Phi_{b,P}$ with Poisson arrivals $\Phi_u$, and in Fig.\ref{VoronoiStations:Lattice} for the case of a square lattice $\Phi_{b,L}$ with Poisson arrivals $\Phi_u$. 
We also provide the reader with a list of symbols in Table \ref{TofS}.

\begin{table}[t!]
\caption{Symbols}
\centering
\begin{tabular}{|c | l |}
\hline
$\Phi_b$ 							& Point Process of transmission nodes $\left\{x_i\right\}$\\
$\Phi_{b,P}$, $\Phi_{b,L}$ 			& Poisson and Lattice position of $\left\{x_i\right\}$\\
$\Phi_u$ 							& Point Process of users marked by object $\left\{(\psi_i,t_i,z_i)\right\}$\\
$\lambda_b$ 						& intensity of transmission nodes [$m^{-2}$]\\
$\lambda_u$ 						& intensity of users [$m^{-2}sec^{-1}$]\\
$A$ 								& planar area\\
$\mathcal{V}(x_i)$					& Voronoi cell of node $x_i$\\
$\mathcal{C}_i$					& coverage cell of node $x_i$\\
$R_b$ 							& radius of coverage\\
$p_m$							& probability of coverage by m nodes\\
$\mathcal{F}$						& object catalogue of size $F$\\
$a_j$							& popularity of object $c_j\in\mathcal{F}$\\
$o$, $u_o$ 						& Typical location and typical user\\
$K$ 								& size of cache memory\\
$\Xi_i(t)$							& inventory of cache on BS $x_i$ at time $t$\\
$\Phi_c$							& Point Process of new content arrivals\\
$\lambda_c$						& intensity of new content arrivals [$obj\ day^{-1}$]\\
$\tau_n$							& lifespan of content $n$\\
$v_n$							& volume (total demands) of content $n$\\
$g_n$							& popularity shape of content $n$\\
\hline
\end{tabular}
\label{TofS}
\end{table}

\section{Multi-LRU under Spatial IRM traffic}
\label{Sec:3_IRM}

The new policies are first evaluated under spatial IRM traffic \cite{FaPrSIAM78}, which has the following properties: 

(i) The $\Phi_u$ is a homogeneous Poisson Point Process (PPP) in both space and time, with intensity $\lambda_u>0$ in $[m^{-2}sec^{-1}]$. Given a planar area $A$, the arrival rate of users in this area, with any request, is equal to $\lambda_u|A|$ in $[sec^{-1}]$. 
All users within the area take their positions independently and uniformly. 

 (ii) The catalogue of available files/objects is fixed over time $\mathcal{F}(t):=\mathcal{F}$, and has finite size $F$. The elements of $\mathcal{F}$ are $\left\{c_1,\ldots,c_F\right\}$. We additionally consider that all objects have the same size, normalised to 1. Cases of unequal size will not be treated in this work, but we can always assume that each file can be divided into chunks of equal size, so the same analysis can still be applied. 
 
 (iii) The probability $a_j$ that a user requests object $c_j\in\mathcal{F}$ (i.e. the object \textit{popularity}) is constant over time, (can be) known, and independent of all past requests. Objects in $\mathcal{F}$ are ordered by popularity: $c_1$ is the most popular, $c_2$ the second most popular and so on. The popularity of $c_j$ is $a_j>0$, and to be consistent with the ordering, we also have $a_1\geq a_2\geq\ldots \geq a_F$. For every popularity distribution it obviously holds,
$\sum_{j=1}^{F}a_j = 1$. Then the marks $z_i$ are i.i.d. random variables distributed as $Z$ with mass function $\left\{a_j\right\}$. 
A consequence is that the users that request object $c_j\in\mathcal{F}$ form a homogeneous space-time PPP with intensity $a_j\lambda_u$ $[m^{-2}sec^{-1}]$ (independent thinning of $\Phi_u$).

Without loss of generality, we will consider (especially in the simulations) that the distribution has a Zipf probability mass function, although the analysis holds for general $\left\{a_j\right\}$. This is motivated by traffic measurements showing that data-object popularity in the WWW follows a power law \cite{HuberWWW99}, \cite{Newman05}. In such case, the probability that a user asks for $c_j$ is equal to 
$a_j  =  D^{-1}j^{-\gamma}$, $j=1,\ldots,F$.
Here, $\gamma$ is the Zipf exponent, often satisfying $\gamma<1$, so that $a_1/a_2=2^{\gamma}<2$. The normalisation factor is equal to $D:=\sum_{j=1}^{F} j^{-\gamma}$.


\textit{Hit performance upper bound:} As mentioned already, the performance measure of the caching policies is the hit probability. We can already provide an upper bound for any POP policy under IRM traffic. The bound requires knowledge over the content popularity and coverage number, like the PBP \cite{BlaGioICC15}. The main idea is that the hit probability of a user covered by $m$ cells is maximised if these $m$ inventories have distinct entries, so that the user has the maximum choice. Hence if the $mK$ most popular objects from the set $\mathcal{F}$ are installed,
\begin{eqnarray}
\label{UBhit}
P_{hit} \leq \sum_{m=1}^M p_m \sum_{j=1}^F a_j \mathbf{1}_{\left\{1\leq j\leq mK\right\}} =  \sum_{m=1}^M p_m \sum_{j=1}^{mK} a_j. 
\end{eqnarray}

\subsection{Che approximation for IRM traffic: Single Cache}

The mathematical analysis of LRU policies is complicated, due to the different inter-arrival times for different content and the update/insertion policy. However, Che et al provided in 2002 \cite{CheApprox02} an analysis and a simple approximation for the single-LRU cache, which results in an analytic formula for the hit probability $P_{hit}$ with excellent fit to simulations. In the following, we explain in short the idea and, after, apply it to the multi-LRU policies.

The approximation is based on the so-called \textit{characteristic time} $T_C$. Given a cache of size $K$ under single-LRU replacement, if at time $t=0$ an arrival of object $c_j$ occurs, then this will be positioned at the MRU place, either due to a. Update, or due to b. Insertion. This object is removed from the cache if at least $K$ different objects arrive, before a new demand for object $c_j$ at time $s_j>0$. The reason is that, each arrival of a new object moves $c_j$ one position away from the MRU and closer to the LRU. 
Che et al approximate the eviction time of an object by a deterministic quantity, equal for all objects to the characteristic time $T_C$. This is found by solving
\begin{eqnarray}
\label{Cheapprox}
\sum_{i=1}^F \mathbb{P}\left(s_i<T_C\right) = K & & (Che\ approximation),
\end{eqnarray}
using a fixed point procedure, where $s_i$ is the first arrival time of object $c_i$, $i\neq j$, after $t=0$. The summation in (\ref{Cheapprox}) is taken over the entire $\mathcal{F}$, which is also part of the approximation. It works well for a large number $F$ of objects, each one of which having a small portion of the popularity. For IRM traffic, the inter-arrival times are exponentially distributed, hence for an area $A$ covered by a single cache, $\mathbb{P}\left(s_i<T_C\right) = 1-e^{-\lambda_u |A|a_i T_C}$. The time-average probability that an object $c_j$ is in the cache is $\mathbb{P}\left(c_j\in\Xi\right) = \mathbb{P}\left(s_j<T_C\right)$, hence
\begin{eqnarray}
\label{Pin}
\mathbb{P}\left(c_j\in\Xi\right) \stackrel{IRM}{=} 1-e^{-\lambda_u |A|a_j T_C}  \stackrel{PASTA}{=} P_{hit}(j).
\end{eqnarray}
The fact that, for IRM traffic $\mathbb{P}\left(c_j\in\Xi\right) = P_{hit}(j)$, is due to the PASTA property of Poisson arrivals. Finally, the approximation for the total hit probability is, 
\begin{eqnarray}
\label{PhitChe}
P_{hit} & = & \sum_{j=1}^F a_j P_{hit}(j).
\end{eqnarray}

%
\subsection{Che-like approximation for multi-LRU}
\label{GenChe}


We will use the approach of Che for the single-LRU, to derive here similar approximations of the multi-LRU cache management policies, for the network model described in the previous section. (To provide more intuition on this general approach, a similar analysis for a network with only two caches is given in the Appendix).

Consider an arrival of user $u_o$ at the Cartesian origin $\psi_o=(0,0)$ at time $t_o=0$, who requests for object $c_j$. This is the \textit{typical user}, who is covered by a number $m_o\geq 0$ of BSs, a realisation of the r.v. $\mathcal{N}_o$ with mass function $\left\{p_m\right\}$. A common characteristic time $T_C$ is assumed for all caches of the network, due to stationarity of all processes. We focus on the cache of a specific $x_i$ among the $m_o$ covering BSs, for which definitely $o\in\mathcal{C}_i$. The probability that user $u_o$ finds the requested content in the cache of $x_i$, is calculated using the following arguments:  There is a previous user $u_{-1}$ requesting for the same object $c_j$, who arrived in an area $\mathcal{S}_{-1}$ (that varies depending on the type of multi-LRU policy). The $u_{-1}$ is covered by $x_i$ definitely (otherwise the user will not influence $\Xi_i$) and possibly some other stations, the total number of which is $\tilde{m}$ (the realisation of another r.v. $\mathcal{N}_{-1}$). Since we know that $u_{-1}$ is at least covered by one station (the $x_i$), the distribution of $\mathcal{N}_{-1}$ has mass function 
\begin{eqnarray}
\label{pmac}
\tilde{p}_{\tilde{m}} = p_{\tilde{m}}({1-p_0})^{-1}, & & \tilde{m}=1,\ldots,M.
\end{eqnarray}

Suppose this user arrived at $t_{-1}^-\in\left[t_o-T_C,t_o\right]$, i.e. within the characteristic time ($t^-$ is the time right before $t$). Then 
the object is found in $\Xi_i(t_o^-)$ at $t_o^-$, if (i) either the object was in $\Xi_i(t_{-1}^-)$ and an update was triggered by $u_{-1}$, or (ii)  the object was not cached in any of the $\tilde{m}$ stations and an insertion in the inventory $\Xi_i$ was triggered. If $m_o>0$ (otherwise, the user is not covered), we write for $ i\in\left\{1,\ldots,m_o\right\}$
\begin{eqnarray}
\label{Phitgen1}
P_{hit,i}(u_o) & = & \mathbb{P}\left(u_{-1}\in(\mathcal{S}_{-1},|t_o-t_{-1}|<T_C,j)\right)\cdot\nonumber\\
& & \cdot \left[\mathbb{P}(c_j\in\Xi_i(t_{-1}^-))+\mathbb{P}(\bigcap_{\ell=1}^{\tilde{m}}\left\{c_j\notin\Xi_\ell(t_{-1}^-)\right\})\right].\nonumber
\end{eqnarray}
For IRM traffic with PASTA, $\mathbb{P}(c_j\in\Xi_i(t_{-1}^-)) = P_{hit,i}(u_{-1})$, and is also independent of the time $t$ and user position $\psi$, hence we can simply write $P_{hit,i}(j)$. Substitution in the above equation gives,
\begin{eqnarray}
\label{Phitgen2}
P_{hit,i}(j) & = & \mathbb{P}\left(u_{-1}\in(\mathcal{S}_{-1},|t_o-t_{-1}|<T_C,j)\right)\cdot\nonumber\\
& & \cdot \left[P_{hit,i}(j) +\sum_{\tilde{m}=1}^{M} \frac{p_{\tilde{m}}}{1-p_o}\mathbb{P}(\bigcap_{\ell=1}^{\tilde{m}}\left\{c_j\notin\Xi_\ell\right\})\right].\nonumber\\
\end{eqnarray}
Solving the above over $P_{hit,i}(j)$ provides an expression for the hit probability of object $c_j$ at the cache of node $x_i$. To find the characteristic time $T_C$ we solve the equation,
\begin{eqnarray}
\label{TCgen}
\sum_{j=1}^F P_{hit,i}(j) = K, & & i\in\left\{1,\ldots,m_o\right\}.
\end{eqnarray}
Finally, the total hit probability is equal to,
\begin{eqnarray}
\label{PhitTOTgen}
P_{hit} 
& = & \sum_{j=1}^Fa_j\sum_{m_o=0}^{M}p_{m_o}\left(1-\mathbb{P}(\bigcap_{\ell=1}^{m_o}\left\{c_j\notin\Xi_\ell\right\})\right).
\end{eqnarray}
We note that $\mathbb{P}(\bigcap_{\ell=1}^{0}\left\{c_j\notin\Xi_\ell\right\}) = 1$, for $m_o=0$, in which case, the user surely misses the content.

The main difficulty when dealing with the general case, is that the hit probability of one cache depends on the hit probability of its neighbours and the neighbours of its neighbours. This is because the coverage area of each node has many sub-areas of multi-coverage by different BS subsets, which makes analysis neither easy, nor exact. 

$\bullet$ {\textbf{multi-LRU-One (Che with CIA)}} 
Only the users falling in the Voronoi cell of a node can trigger an action of a. Update or b. Insertion at the cache of that node as long as they are covered. (Here we analyse this version for the Update phase, but other policy variations are possible without significant performance change). Then $\mathcal{S}_o=\mathcal{S}_{-1}=\mathcal{V}(x_i)$ in (\ref{Phitgen1}). The coverage cell can be smaller than the Voronoi cell, in which case only the users falling in the intersection of the two trigger cache actions. To avoid dealing with these special cases, we consider coverage cells which fully cover the related Voronoi cells, that is $|\mathcal{C}_i|>|\mathcal{V}_i|$, $\forall i$.


There are the unknown probabilities $\mathbb{P}(\bigcap_{\ell=1}^{\tilde{m}}\left\{c_j\notin\Xi_\ell\right\})$ and  $\mathbb{P}(\bigcap_{\ell=1}^{m_o}\left\{c_j\notin\Xi_\ell\right\})$ that need to be calculated. Instead of directly trying to find a solution, we use a \textit{Cache Independence Approximation} (CIA). Based on this, each cache performs single-LRU for the users that arrive within its Voronoi cell. The idea is that, since only the users in the Voronoi cell change the inventory of the related cache, the influence of the neighbouring stations' traffic on the inventory of $x_i$ should be small. Then in (\ref{Phitgen1}) we forget the rest $\tilde{m}-1$ nodes and we replace 
\begin{eqnarray}
\label{CIA1}
\mathbb{P}(\bigcap_{\ell=1}^{\tilde{m}}\left\{c_j\notin\Xi_\ell\right\}) \approx \mathbb{P}(c_j\notin\Xi_i), & (CIA_1).
\end{eqnarray}
Furthermore, the independence due to the CIA, has the result that, when the user is covered by $m_o$ stations, her/his hit probability is simply the product of hit probabilities of all these stations. The fact that the Voronoi cells of different stations do not overlap is further in favour of the approximation. Then, in (\ref{PhitTOTgen})
\begin{eqnarray}
\label{CIA2}
\mathbb{P}(\bigcap_{\ell=1}^{m_o}\left\{c_j\notin\Xi_\ell\right\}) \approx (\mathbb{P}(c_j\notin\Xi_i))^{m_o}, & (CIA_2).
\end{eqnarray}
From the above, the hit probability of each object in $\Xi_i$ is,
\begin{eqnarray}
\label{Phitgen2one}
P_{hit,i}(j) & = & \mathbb{P}\left(u_{-1}\in(\mathcal{S}_{-1}\in\mathcal{V}(x_i),|t_o-t_{-1}|<T_C,j)\right)\cdot\nonumber\\
& &  \cdot\left[P_{hit,i}(j) +\mathbb{P}(c_j\notin\Xi_i)\right]\nonumber\\
& \stackrel{IRM}{=}& 1- e^{-a_j\lambda_u|\mathcal{V}|T_C}, \ i\in\left\{1,\ldots,m_o\right\}.
\end{eqnarray}
We used the fact that for IRM $P_{hit,i}(j) = 1- \mathbb{P}(c_j\notin\Xi_i)$.
The characteristic time is found by solving the equation
\begin{eqnarray}
\label{TConegen}
\sum_{j=1}^F (1- e^{-a_j\lambda_u|\mathcal{V}|T_C}) =K.
\end{eqnarray}
The total hit probability, based on CIA, is, 
\begin{eqnarray}
\label{PhitTOTgenone}
P_{hit} & = & \sum_{j=1}^Fa_j\sum_{m_o=0}^{M}p_{m_o}\left(1-\mathbb{P}(c_j\notin\Xi_i)^{m_o}\right)\nonumber\\
& \stackrel{(\ref{Phitgen2one})}{=} & \sum_{j=1}^Fa_j\sum_{m_o=0}^{M}p_{m_o}\left(1-e^{-a_j\lambda_um_o|\mathcal{V}|T_C}\right).
\end{eqnarray}
Special case: For the PPP model of node positions, the Voronoi cell size is a random variable. We can use for simplicity of the expression the average size of a Voronoi cell, equal to $|\mathcal{V}|=\lambda_b^{-1}$, \cite{BacBlaVol1}. In the Boolean coverage model, $|\mathcal{C}|=\pi R_b^2$. 


$\bullet$ {\textbf{multi-LRU-All (Che with CSA)} }
In this case, users falling on any point inside the coverage cell of $x_i$ can trigger an action of update and insertion at its cache inventory $\Xi_i$. This means that $\mathcal{S}_{o}=\mathcal{S}_{-1}=\mathcal{C}_i$, for the hit probability expression in (\ref{Phitgen1}). 

Again, the unknown probabilities $\mathbb{P}(\bigcap_{\ell=1}^{\tilde{m}}\left\{c_j\notin\Xi_\ell\right\})$ and $\mathbb{P}(\bigcap_{\ell=1}^{m_o}\left\{c_j\notin\Xi_\ell\right\})$ need to be calculated. 
In this case, we use a different approximation, the \textit{Cache Similarity Approximation} (CSA), which states that inventories of neighbouring caches have the same content. This is motivated by the fact that new content is simultaneously installed in all caches of nodes covering a user, when the user triggers insertion. The approximation is better, the larger the cache size $K$, because for large memories it takes more time for an object to be evicted after its insertion and similar content stays in all inventories. Then in (\ref{Phitgen1}),
\begin{eqnarray}
\label{CSA1}
\mathbb{P}(\bigcap_{\ell=1}^{\tilde{m}}\left\{c_j\notin\Xi_\ell \right\}) \approx  \mathbb{P}(\left\{c_j\notin\Xi_i\right\}), & (CSA_1).
\end{eqnarray}
Interestingly, $CSA_1$ and $CIA_1$ give the same expression. However, in multi-LRU-All, we do not assume independence, but rather similarity. Then, since neighbouring caches have the same content, the total miss probability when a set of $m_o$ stations cover user $u_o$ is equal to the probability that no user with the same demand arrives within the total area of coverage during the characteristic time $T_C$ (otherwise the content is definitely in all caches, either because of a. Update or b. Insertion. Then, for IRM traffic,
\begin{eqnarray}
\label{Pmissmoall}
\mathbb{P}(\bigcap_{\ell=1}^{m_o}\left\{c_j\notin\Xi_\ell\right\}) \approx e^{-a_j\lambda_u|\mathcal{A}_{m_o}|T_C}, & (CSA_2).
\end{eqnarray}
In the above, the total area of coverage from the $m_o$ stations is denoted by $\mathcal{A}_{m_o}$ and its surface is equal to,
\begin{eqnarray}
\label{SurfaceM}
\left|\mathcal{A}_{m_o}\right| & = & \left|\bigcup_{\ell=1}^{m_o}\mathcal{C}_i\right|, \ \  m_o=0,\ldots,M.
\end{eqnarray}
It holds $|\mathcal{A}_0|=0$, for $m_o=0$. For the Boolean model $|\mathcal{A}_{1}| = \left|\mathcal{C}_1\right| = \pi R_b^2$, while the surface of $\mathcal{A}_{m_o}$ is a superposition of $m_o$ overlapping discs with equal radius $R_b$.

The hit probability of each object in $\Xi_i$ is found by using CSA in (\ref{Phitgen1}), and we get
\begin{eqnarray}
\label{Phitgen2all}
P_{hit,i}(j) & = & \mathbb{P}\left(u_{-1}\in(\mathcal{S}_{-1}\in\mathcal{C}_i,|t_o-t_{-1}|<T_C,j)\right)\cdot\nonumber\\
& &  \cdot \left[P_{hit,i}(j) +\mathbb{P}(\left\{c_j\notin\Xi_i\right\})\right]\nonumber\\
& \stackrel{IRM}{=}& 1- e^{-a_j\lambda_u|\mathcal{C}|T_C}.
\end{eqnarray}
We used the fact that for IRM $P_{hit,i}(j) = 1- \mathbb{P}(c_j\notin\Xi_i)$. For the characteristic time, we solve the equation
\begin{eqnarray}
\label{TCallgen}
\sum_{j=1}^F (1- e^{-a_j\lambda_u|\mathcal{C}|T_C}) & = & K.
\end{eqnarray}
The total hit probability, based on CSA, is
\begin{eqnarray}
\label{PhitTOTgenall}
P_{hit} & \stackrel{(\ref{Pmissmoall})}{=} & \sum_{j=1}^Fa_j\sum_{m_o=0}^{M}p_{m_o}\left(1-e^{-a_j\lambda_u|\mathcal{A}_{m_o}|T_C} \right).
\end{eqnarray}
The difficulty in calculating the approximate hit probability for multi-LRU-All with the above formulas, is to obtain exact values for the total surface $|\mathcal{A}_{m_o}|$. For the PPP special case, this surface is also a random variable, that we can approximate by its mean value for simplicity of the expressions. A method to approximate these quantities is given in V.A. (We refer, again, the reader to the Appendix for the two-cache network example.)

%
%
\section{Multi-LRU under temporal locality traffic}
 \label{Sec:4_Temp}

Although the IRM offers tractability, it is not enough to describe real traffic aspects. In real networks new objects (never requested before) appear, while older ones become obsolete after some time. Furthermore, the popularity of a content does not remain constant but varies over time, and there is dependence between requests of the same object within some time window. All these characteristics are described by the term 
\textit{temporal locality} \cite{BeMASCOTS00}, \cite{TraversoTranMult15}, \cite{OlmosTEMPO14} (see also \cite{BroSceWa12} for space-locality). Generators of such traffic have been proposed in the literature \cite{Almeida96}. A so-called Shot Noise Model (SNM) is presented in \cite{TraversoTranMult15}, \cite{LeoINFO15}, which we make use here as a basis of our own traffic model. Under SNM the demand process is a superposition of independent point processes (not necessarily homogeneous), one for each content.



A detailed description of the SNM variation in this work follows. $\mathcal{F}(t)$ is the catalogue (set) of active objects at time $t$, with cardinality $F(t) := |\mathcal{F}(t)|$. The evolution of the catalogue size is a random process. We assume that the arrival of a new object $c_n$ coincides with the time of its first request $t_{n}$. The time instants of these first requests (arrivals) are modelled as a homogeneous PPP $\Phi_c$ on $\mathbb{R}$ with intensity $\lambda_c>0$ [$\frac{objects}{unit-time}$]. Unit-time can be e.g. $1\ day$. 

A pair of r.v.'s is related to each content as an independent mark on the arrival process: (a) The first r.v. denoted by $T_{n}$ is the $n$'th content's \textit{lifespan}, which gives the length of time 
period during which it is requested by users, and after the period's end it becomes obsolete. We could allow for the realisation $\tau_{n}$ to take infinite values but in such option the size of the catalogue would grow indefinitely, unless the popularity of different objects tends to zero fast enough. We let here $\tau_{n}<\infty$ so that the catalogue size $F(t)$ fluctuates over time and remains finite. The time interval of an object is $\Delta t_n:=[t_{n}, t_{n}+\tau_{n})$. (b) The second r.v. attached to the object $c_n$ is the \textit{volume} $V_{n}$ (with realisation $v_n$) i.e. the total number of requests during its life. The pair of values $(\tau_n,v_{n})$ per object is chosen independently of other objects and in the general case should be drawn from a joint probability distribution with a given density $f_{(T, V)}(\tau,v)$, where $T$ and $V$ are the generic variables. In general the two variables are dependent.

To simplify the traffic model it is assumed here that $T$ and $V$ are independent of each other, i.e. $f_{(T, V)}(\tau,v) = f_T(\tau)f_{V}(v)$. This simplification has no obvious impact on the performance of the caching policies. Both lifespan and volume follow a Power-law, i.e both $T$ and $V$ are Pareto distributed \cite{Newman05}. The Pareto distribution in both cases has parameter $\beta>1$ (for the expected value to be finite), and its p.d.f. is given by (here for $V$) $f_V(v)=\frac{{\beta}V_{\min}^{\beta}}{{v}^{\beta+1}}$. Its expected value depends on the values of $\beta$ and $V_{\min}$ through the expression $\mathbb{E}[V] = \frac{\beta V_{\min}}{\beta-1}$. 
To guarantee $V\in\mathbb{N}_+$ for the samples, we choose $V_{\min}=0.5$ and we round to get discrete values. Sampling from a Pareto distribution, generates Zipf-like distributed sizes of objects due to the Power-law behaviour.

Having sampled $(\tau_n,v_n)$ for a specific object arriving at $t_n$, it remains to determine how these $v_n$ requests are positioned within $\Delta t_n$. To include additional attributes of temporal locality in the traffic model, we let requests be distributed according to a finite point process (given $V<\infty$) and more specifically a (non-homogeneous) binomial point process (BPP) $\Psi_n$ on $\mathbb{R}^{v_n-1}$ with density function $g_n(t,t_n,\tau_n)$ over $t$,
\begin{equation}
\label{PsiBi}
 \Psi_n \sim \mathrm{Binomial}(\Delta t_n,v_n-1,g_n(t,t_n,\tau_n)). 
 \end{equation}
 We randomly position only $v_n-1$ requests, because the first request always coincides with the time of content arrival $t_n$. The choice of the $\mathrm{Binomial}$ distribution further implies that requests take position independently of each other. Since $g_n(t,t_n,\tau_n)$ describes how each of the $v_n-1$ requests is distributed within $\Delta t_n$ according to the function's \textit{shape}, the higher the value of $g_n$ for some $t$, the more probable it is that a request will appear at that point. The popularity shape is an important aspect of the model. For some $c_n$ it holds
\begin{equation}
\label{eqg}
g_n(t,t_n,\tau_n)=0 \quad \mathrm{for} \quad t\notin\Delta t_n,
\end{equation}
and
\begin{equation} \label{volume_pop}
\int_{t_n}^{ t_n+\tau_n} g_n(t,t_n,\tau_n) dt=1.
\end{equation}

When the requests follow a homogeneous BPP for some $c_n$ with a given $\Delta t_n$, the shape function is \textit{uniform} and takes the expression $g_n(t,t_n,\tau_n)=\tau_n^{-1}\mathbf{1}_{\left\{t\in\Delta t_n\right\}}$. For further shape options we refer the reader to \cite{Richier14}. In this reference work, for finite volume per object three shapes are proposed, namely (i) the \textit{logistic}, (ii) the \textit{Gompertz}, and (iii) the \textit{negative exponential}\footnote{
By assuming that the lifespan of every content ends when it reaches its $1-\varepsilon$ of its total views $v_i$, $\tau_i$ can be mapped to the curve parameter $\lambda$ in \cite{Richier14}, which parametrises the speed of change of a content's popularity. In this paper we chose $\varepsilon=0.02$.}. Applying this to our traffic generator, when a new object arrives it is assigned a shape of index $k$ with probability $a_k$, the exact value of which is a tuneable parameter. 
%

The spatial (geographic) aspect of traffic plays an important role in influencing the performance of the policies studied here. In our work requests are uniformly positioned on a finite 2D plane. However, the traffic model can be easily extended to incorporate spatial locality. 

Based on the above description, characteristic quantities of the generated traffic can be derived.

- \textit{Mean Catalogue Size $\mathbb{E}[\mathcal{F}(t)]$.} Because of the stationarity of the arrival PPP
the expected number of active contents (hence catalogue size) does not depend on time $t\in\mathbb{R}_+$,
\begin{eqnarray}
\label{EF}
\mathbb{E}[\mathcal{F}(t)] & = &  \lambda_c\mathbb{P}(V>1)\mathbb{E}[T].
\end{eqnarray}

- \textit{Mean Total Number of Requests within $[0,t]$} [$days$],
\begin{eqnarray}
\label{Nreq}
N_{req}([0,t]) & = & t\lambda_c\mathbb{E}[V].
\end{eqnarray}

- \textit{Memory-to-Mean-Catalogue-Size-Ratio (MMCSR)} where we omit $\mathbb{P}(V>1)$, which is just a scaling constant,
\begin{eqnarray}
\label{CCSR}
\rho & := & \frac{K}{ \lambda_c\mathbb{E}[T]}.
\end{eqnarray}

\textit{Hit Upper Bound: } 
Similarly to the IRM traffic, we can derive a (numerical) upper bound for the POP policies under traffic with temporal locality. We consider a scenario where popularities are estimated periodically. Specifically, at time instants $t_n=n\Delta t_{up}$, $n\in\mathbb{Z}$, the caches are updated by some POP policy, using the estimated popularities during the time interval $[t_n-\Delta t_{es},t_n)$, i.e. $\Delta t_{es}$ is the window of observation. Let $\mathcal{F}_{mK}(t_n)$ be the set of the $mK$ most requested objects in $[t_n-\Delta t_{es},t_n)$. Then the upper bound within the time interval $[t_n,t_n+\Delta t_{up})$ is equal to,
\begin{eqnarray}
\label{POPub}
P_{hit}^{(POP)}[t_n,t_n+\Delta t_{up})\leq\sum_{m=1}^{\infty}p_m\mathbb{P}(c\in\mathcal{F}_{mK}(t_n)).
\end{eqnarray}
The more often the algorithm updates the caches the better performance it achieves. But, considering that a cache update will use backhaul and computational resources by the controller, $\Delta t_{up}$ cannot be too small. We fix $\Delta t_{up}=1$ $day$, i.e. the caching policy runs every night when the request load is low \cite{GolrezaeiINFOCOM12}. As far as $\Delta t_{es}$ is concerned there is a "crisp" optimal choice. If it is too big, $\mathcal{F}_{mK}(t_n)$ will possibly include outdated objects. On the other hand, small $\Delta t_{es}$ can result in excluding even the most popular objects from $\mathcal{F}_{mK}(t_n)$ because they have not been sufficiently requested. The bound in (\ref{POPub}) can be evaluated by Monte Carlo simulations. The optimum $\Delta t_{es}$ [$days$] can be numerically determined.

%
%
 \section{Simulation and Comparison}
 \label{Sec:5_Simul}



For the simulations, BSs are placed within a rectangular window of size $A \times B = 12\times 12$ $[km^2]$. After choosing the BS intensity $\lambda_b=0.5$ $[km^{-2}]$, their positions are chosen based on the type of network we want to analyse (PPP or Lattice). For PPP, a Poisson number of stations is simulated in each realisation and their positions are set uniformly inside the window. In the case of a Lattice network, the stations are put on a square grid with distance $\eta=1/\sqrt{\lambda_b} = 1.4142$ $[km]$ from each other. In both types of networks, the average Voronoi size $|\mathcal{V}|=\lambda_b^{-1}$ (see \cite{BacBlaVol1}).

We evaluate a Boolean coverage model so that every station covers a disc of radius $R_b\in\left[0.5, 3\right]$ $[km]$ with surface $|\mathcal{C}|=\pi R_b^2$. The larger the radius the stronger the multi-coverage effects. The magnitude of coverage overlap can be described by the expected number of BSs covering a planar point, $\overline{N_{BS}}=\mathbb{E}[{\mbox{Number of covering stations}}] = \sum\limits_{m=1}^{\infty}m p_m$, where the $p_m$ are the coverage number probabilities for $m$ stations, whose values depend on the node placement and coverage model. For the Boolean PPP case, the probabilities $\left\{p_m\right\}$ correspond to a Poisson r.v. with parameter $\nu:=\lambda_b\pi R_b^2$ (see \cite{BacBlaVol1}). For the Boolean Lattice case, these are found by Monte Carlo simulations. Given the intensity, $\lambda_b = 0.5$, there is a mapping from the Boolean radius $R_b$ to the number $\overline{N_{BS}}$, some values of which are given in Table \ref{tab:Radius_Nexp}.
\begin{table}[h]

\caption{$R_b$  to $\overline{N_{BS}}$ mapping: Boolean PPP and Lattice ($\lambda_b=0.5$ $km^{-2}$).}
\centering
\begin{tabular}{|c|c|c|}

\hline
Radius ($R_b$) $[km]$	& PPP ($\overline{N_{BS}}$) & 	Lattice ($\overline{N_{BS}}$) \\
\hline
\hline
0.8 	& 1 & 1.06\\
\hline
1.13 & 2 & 2.12\\
\hline
1.38 & 3 & 3.22\\
\hline
1.60 & 4 & 4.21\\
\hline
1.78 & 5 & 5.32\\
\hline
1.95 & 6 & 6.42\\
\hline
2.11 & 7 & 7.43\\
\hline
2.26 & 8 & 8.44\\
\hline
\end{tabular}
\label{tab:Radius_Nexp}
\end{table}

\begin{figure}[t!]   
\centering  
\label{multiLRUoneallVerif}
\subfigure[multi-LRU-One]{
	  \centering
	         \includegraphics[trim=1cm 0.9cm 1cm 1.4cm, clip, width=0.30\textwidth]{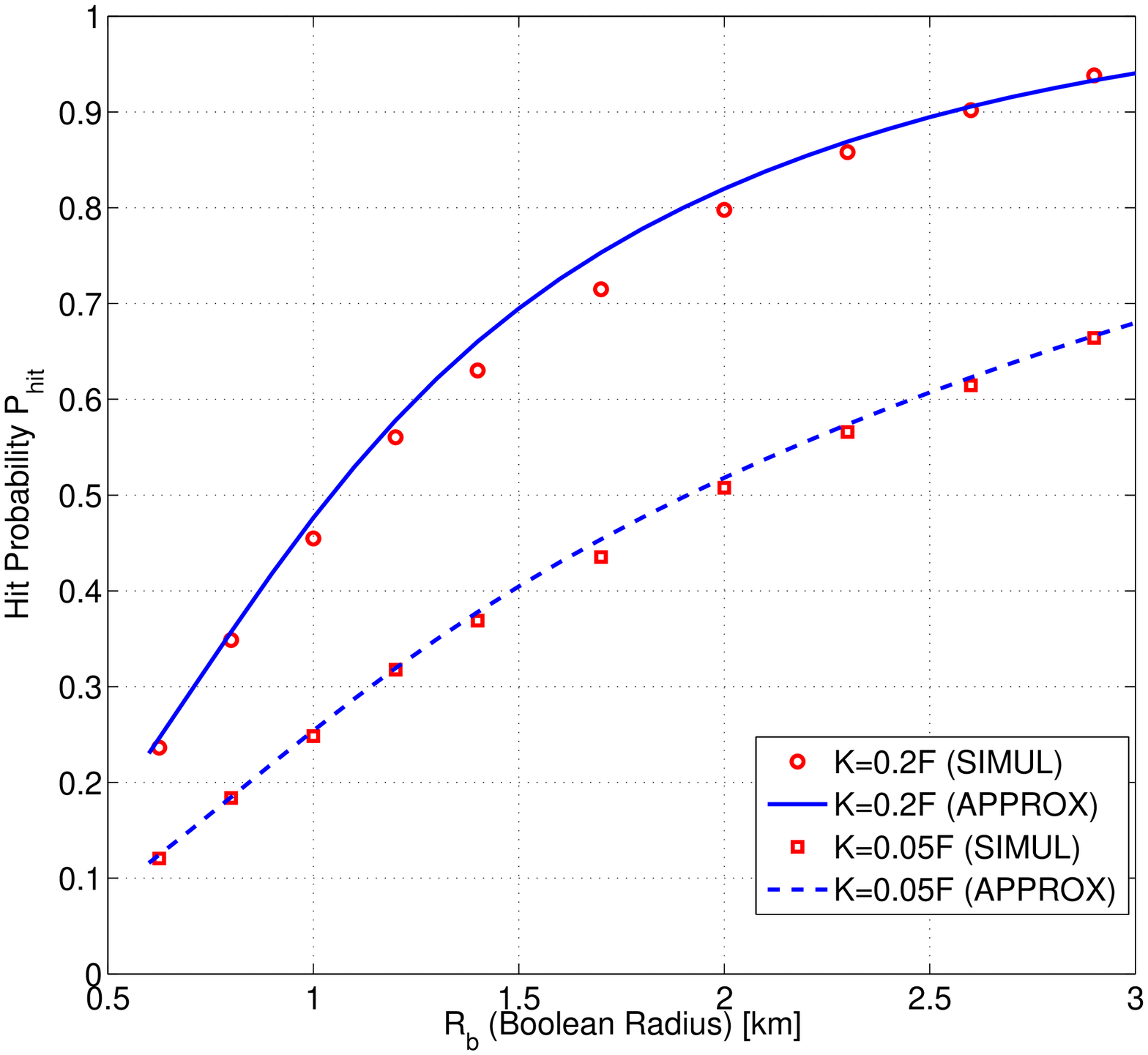}
           \label{Verif:ONE}
           }
           
           \subfigure[multi-LRU-All]{
	   \centering  
	         \includegraphics[trim=1cm 0.9cm 1cm 1.4cm, clip, width=0.30\textwidth]{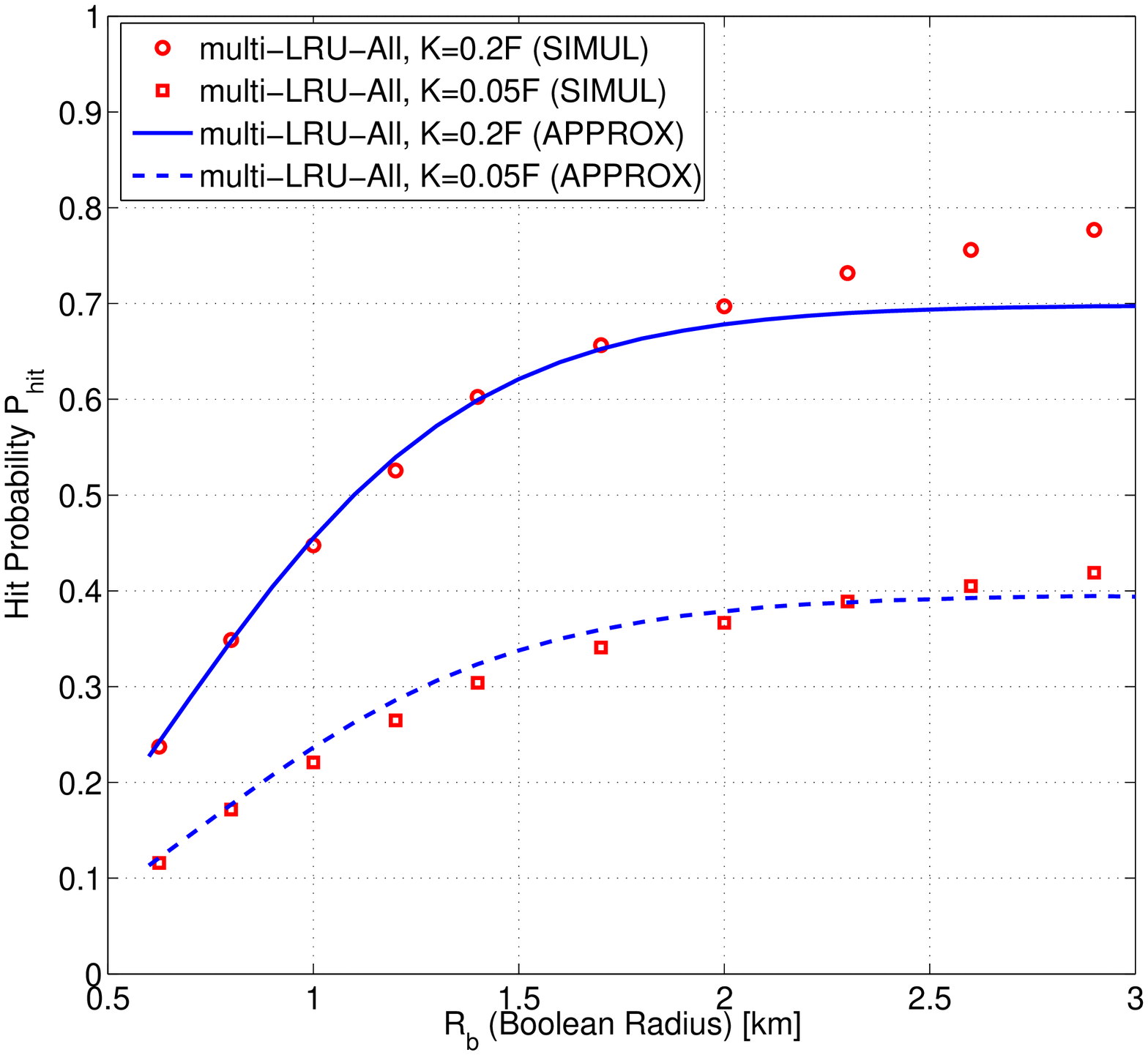}
	   \label{Verif:ALL}
           }
         \caption{Verification of the approximations for the two multi-LRU policies.}
\end{figure}

\subsection{IRM traffic}
Following the spatial IRM traffic model for the request arrivals, we consider a homogeneous space-time PPP with intensity $\lambda_u=0.023$ $[m^{-2}sec^{-1}]$, which is approximately equal to 80 $[m^{-2}/hour^{-1}]$ requests - a reasonable value for a busy corner in a city. 

The content mark for each request is independently chosen from a catalogue of size $F=10,000$ objects. The popularities of these objects follow a Zipf distribution with parameter $\gamma=0.78$  (unless otherwise stated). A cache memory of capacity $K$ objects is considered available on each BS. The size $K$ is defined as a proportion of the catalogue size, i.e. $K/F=\rho$, where  $\alpha$  is called the \textit{Memory-to-Catalogue-Size Ratio (MCSR)}. (We use different notation than the MMCSR $\rho$ for temporal locality in (\ref{CCSR})).


When a user is covered by a station with the requested content in memory, the demand is considered a hit. At the end of the simulation of a large number of realisations for the BS and request point processes (this number chosen over 10,000) the total hit probability is approximated by the frequency of hits (number of hits over number of requests). In the simulations we take considerations over issues related to edge effects that arise from the finite window size.


\subsubsection{Verification of the approximations}
To verify the validity of the proposed approximations, we compare the analytic formulas derived in Section \ref{GenChe} with the hit probability from simulations, for the Boolean PPP coverage model. 
For the memory size $K$, we consider two cases, (a) $\alpha=0.05$, hence $K = 500$ objects, and (b) $\alpha=0.2$, hence $K=2000$ objects. 

$\bullet$ \textbf{multi-LRU-One}: The total hit probability is evaluated numerically using (\ref{PhitTOTgenone}). The characteristic time per cache is found by solving (\ref{TConegen}) by a fixed point method, where the individual hit probability of each object is given in (\ref{Phitgen2one}). To guarantee that $|\mathcal{C}|>|\mathcal{V}|$, we need that $\pi R_b^2>\lambda_b^{-1}$ $\Rightarrow$ $R_b>(\pi\lambda_b)^{-{1/2}}=0.4$. Since $R_b>0.6$ in the evaluation, the condition should be satisfied. The comparison between approximate hit probability and simulations are shown in Fig. \ref{Verif:ONE}. The curves exhibit a very good match. The evaluation shows that the independence approximation (CIA) works very well in this general model with PPs. 

$\bullet$ \textbf{multi-LRU-All}: The total hit probability is evaluated numerically using (\ref{PhitTOTgenall}). The characteristic time per cache is found by solving (\ref{TCallgen}) using a fixed point method, where the individual hit probability of each object is given in (\ref{Phitgen2all}). 

We provide a method to estimate the surfaces $|\mathcal{A}_{m_o}|$, $m_o=1,\ldots,M$ for the Boolean/PPP case: A user $u_o$ has a distance $R_{d,i}$ from each one of the $m_o$ nodes $x_i$ that cover her/him. These distances are realisations of a random variable, whose expected value can be found equal to $\mathbb{E}[R_d]= 2R_b/3$, i.e. the user lies in expectation at $2R_b/3$ away from the center of each covering disc. Then we have:

i) The coverage cell size (for the Boolean model) is the disc surface, equal to $|\mathcal{A}_1|=|\mathcal{C}|=\pi R_b^2$.

ii) When $M\rightarrow\infty$, a disc having center the user $u_o$ and radius $R_M$ $=$ $R_b+\mathbb{E}[R_d]$ $=$ $5R_b/3$ is (due to randomness of node positions) fully covered. So $|\mathcal{A}_M|= |\mathcal{C}|(5/3)^2$.

iii) For intermediate cases $1<m_o<M$, the surface should be somewhere between the two extremes, and obviously the surface $|\mathcal{A}_{m_o}|$ should be monotone increasing with $m_o$. We also expect that  for low $m_o$, the total area $|\mathcal{A}_M|$ will be filling fast, whereas for larger ones, the change in surface should be small. For this we can use a function with exponential decrease for large $m_o$, such as
\begin{eqnarray}
\label{Amo}
|\mathcal{A}_{m_o}| = |\mathcal{A}_M|(1-e^{-m_o\delta}), & \delta = -\ln(1-\frac{|\mathcal{A}_1|}{|\mathcal{A}_M|}).
\end{eqnarray}
The comparison between approximate hit probability and simulations are shown in Fig. \ref{Verif:ALL}. The approximation and simulation curves seem to closely follow one another. For large values of the radius, the approximation curves seem to diverge from the simulations. This should be less a failure of the CSA approximation (which is shown to be accurate for the two-cache network in the Appendix), but more possibly a failure of the above method to approximate well the surfaces $|\mathcal{A}_{m_o}|$. More accurate values of $|\mathcal{A}_{m_o}|$ should exhibit a better fit.

%
\subsubsection{Comparison of policies}

\begin{figure*}
\centering
\label{ALLirm}
    \subfigure[Hit Probability PPP/Boolean, $\alpha=1\%$.]{
    \centering
        \includegraphics[trim=1.1cm 0.5cm 1.2cm 0.5cm, clip, width=0.315\textwidth]{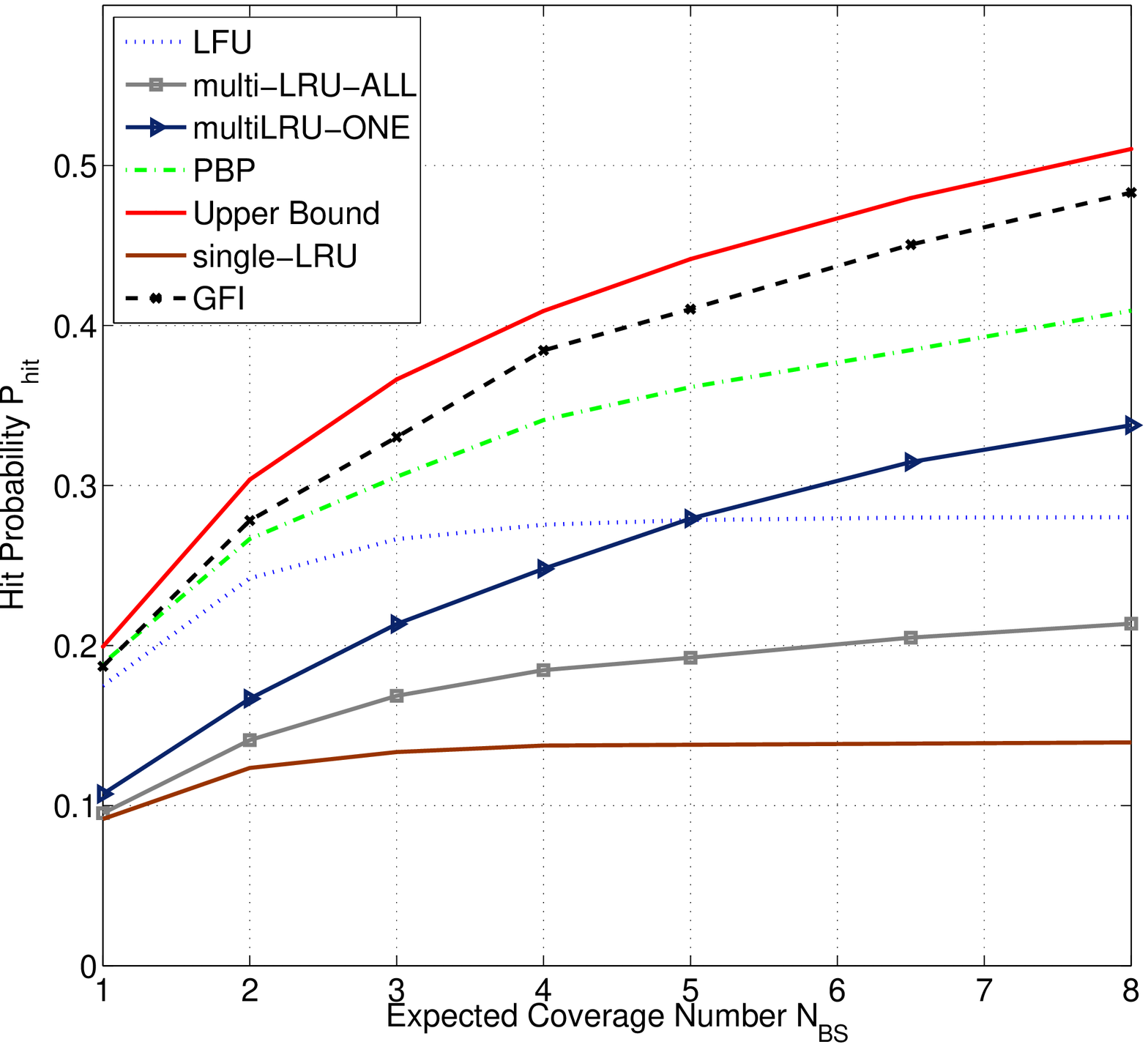}
       \label{fig:PoissonK/F=1}
        }
    \subfigure[Hit Probability Lattice/Boolean, $\alpha=1\%$.]{
        \includegraphics[trim=1.1cm 0.5cm 1.2cm 0.5cm, clip, width=0.315\textwidth]{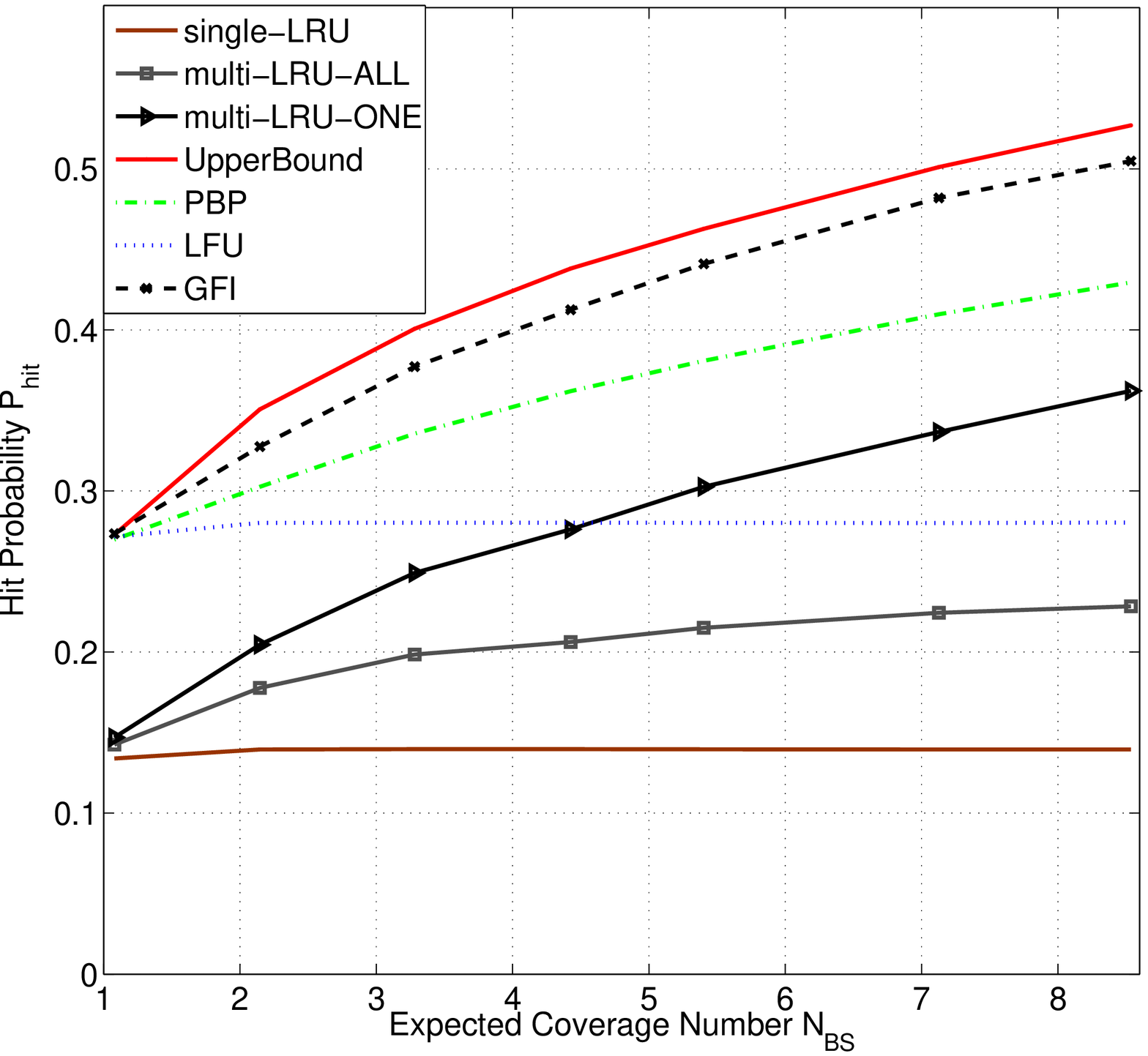}
        \label{fig:LatticeK/F=1}
        }
    \subfigure[Hit Probability Lattice/Boolean, $\alpha=5\%$.]{
        \includegraphics[trim=1.cm 0.5cm 1.2cm 0.5cm, clip, width=0.315\textwidth]{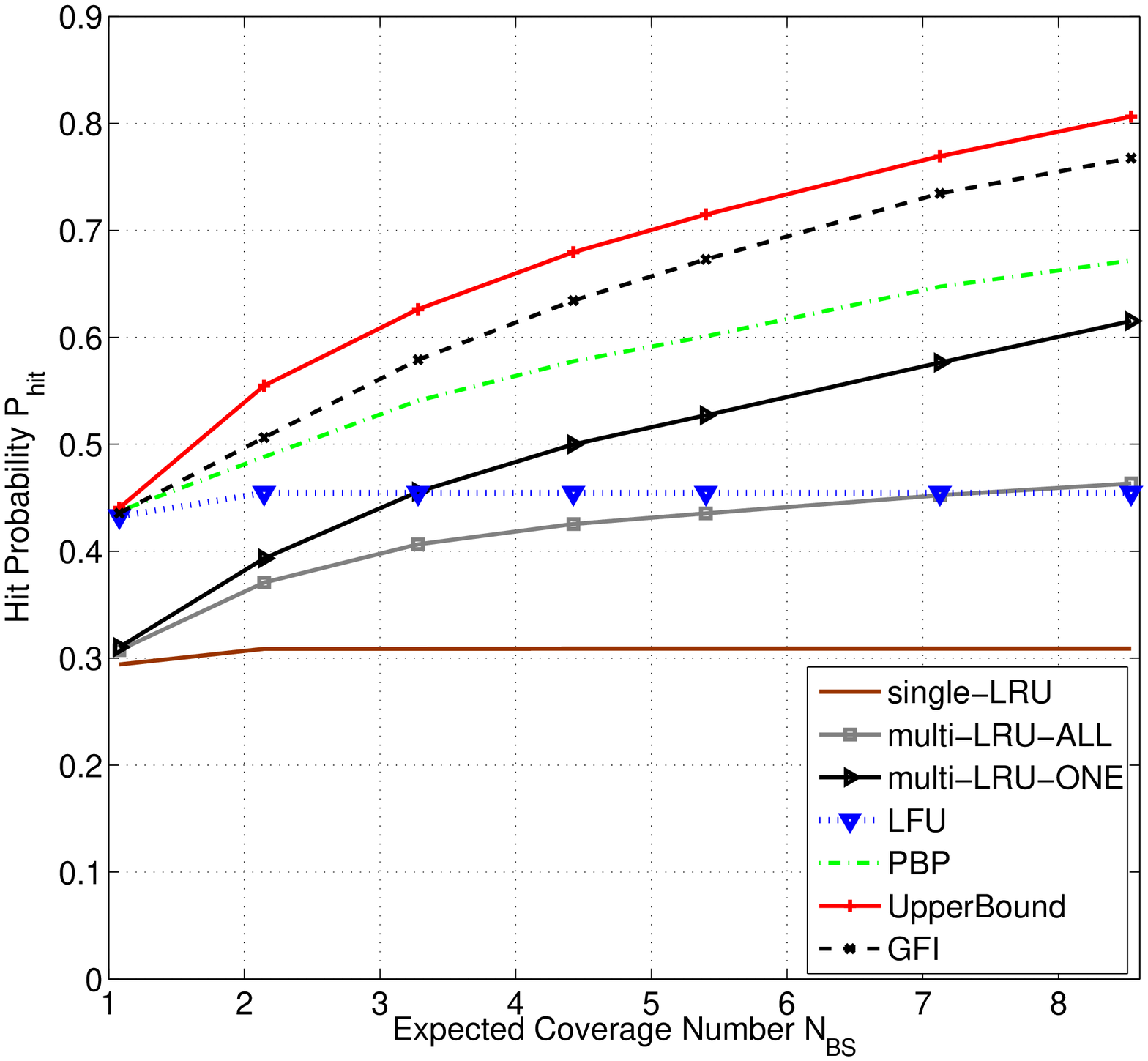}
        \label{fig:LatticeK/F=5}
        }     
        
        \subfigure[q-multi-LRU-All, Boolean/PPP, $\alpha=1\%$.]{
    \centering
        \includegraphics[trim=1.1cm 0.5cm 1.2cm 0.5cm, clip, width=0.315\textwidth]{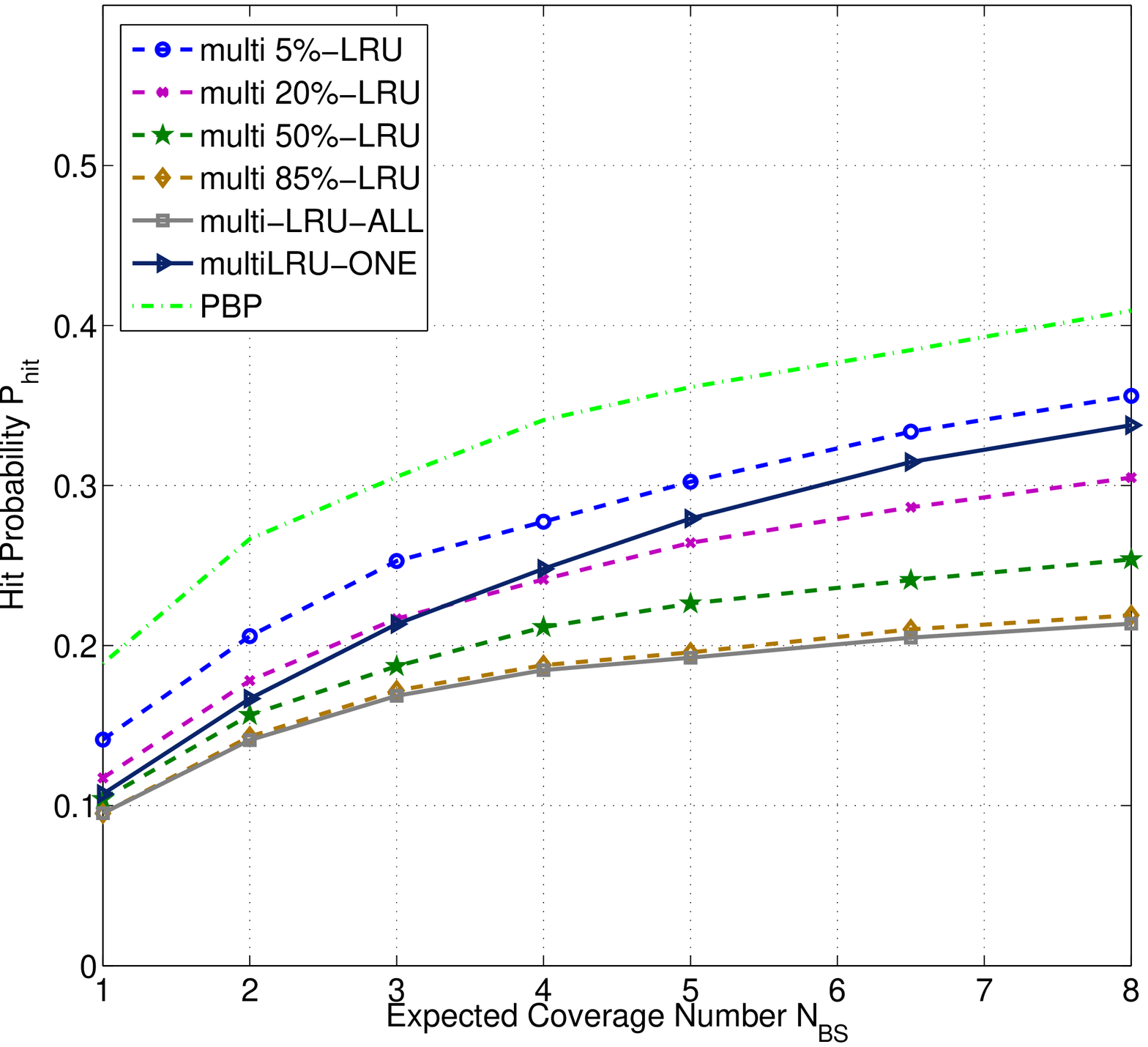}
       \label{fig:PoissonK/F=1_qLRU}
        }
    \subfigure[Hit Probability over $\gamma$, PPP, $\alpha=1\%$.]{
        \includegraphics[trim=1.2cm 0.5cm 1.2cm 0.5cm, clip, width=0.315\textwidth]{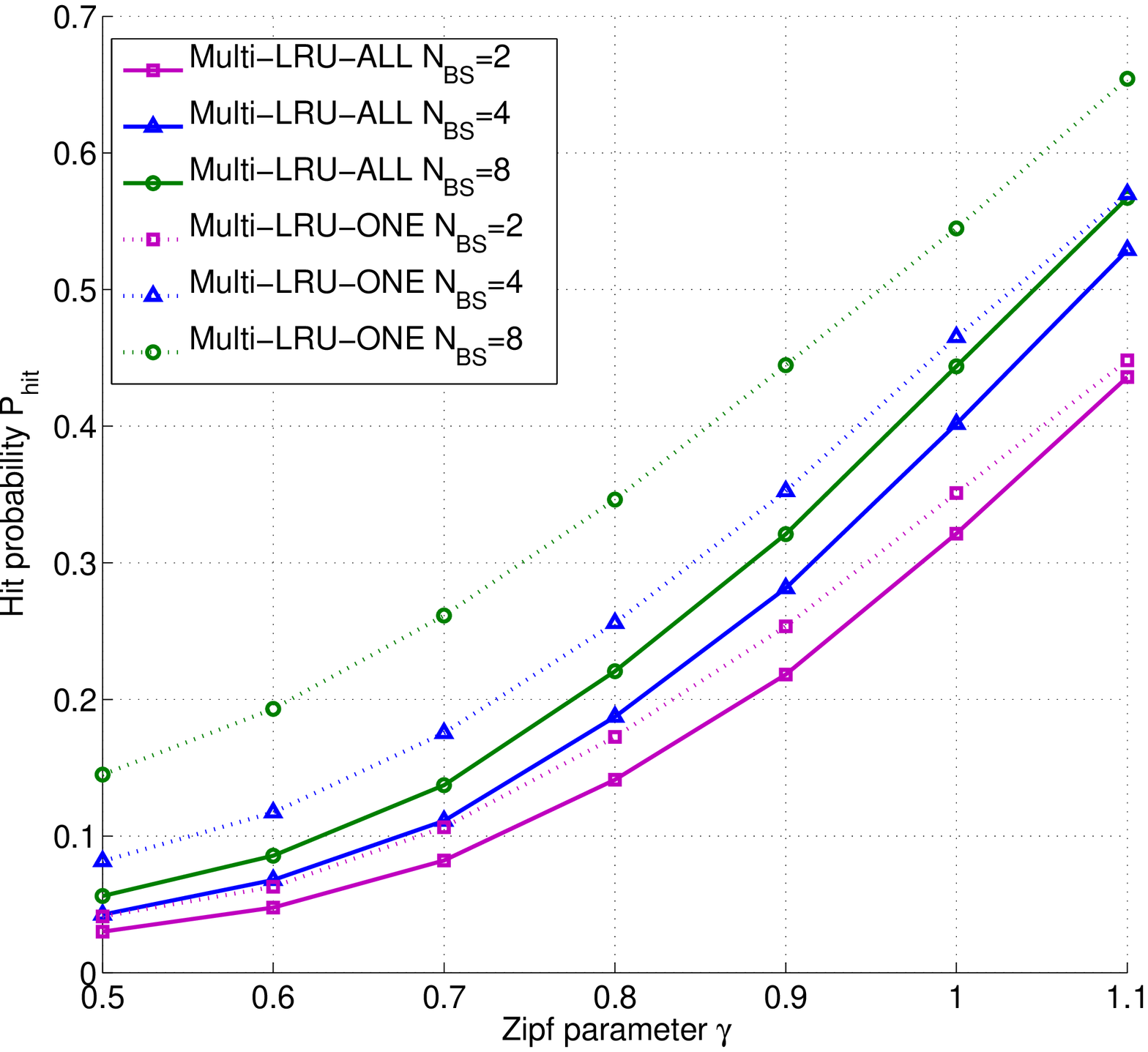}
        \label{fig:VaryGamma_K/F=1}
        }
    \subfigure[Hit Probability over $\rho$, PPP, $\gamma=0.78$.]{
        \includegraphics[trim=1.1cm 0.5cm 1.2cm 0.5cm, clip, width=0.315\textwidth]{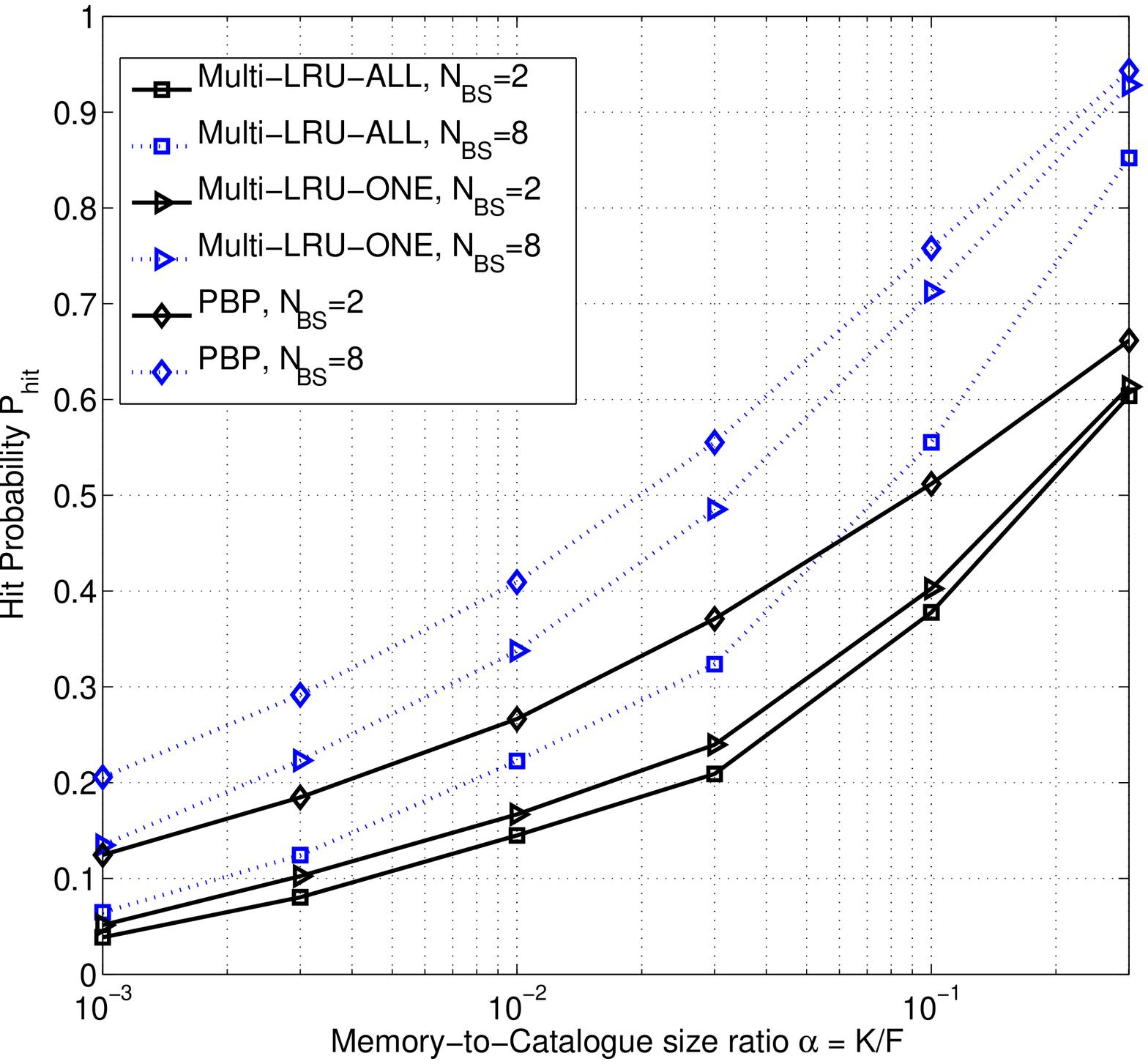}
        \label{fig:VaryK/F}
        }   
    \caption{IRM traffic. Hit Probability evaluation of the two multi-LRU policies and comparison with existing POP and POQ policies.}\label{fig:animals}
\end{figure*}

\textit{Hit versus Coverage Number}. 

In Fig. \ref{fig:PoissonK/F=1}, \ref{fig:LatticeK/F=1} and \ref{fig:LatticeK/F=5} we evaluate the hit probabilities of the proposed \textit{multi-LRU} policies over the expected number of covering stations. In the simulations the radius of the Boolean model is increased from $R_b=0.6$ to $2.25$. The radius is mapped to the expected coverage number, as in Table \ref{tab:Radius_Nexp}. In Fig. \ref{fig:PoissonK/F=1} transmission nodes are positioned as a PPP, while in Fig. \ref{fig:LatticeK/F=1}, \ref{fig:LatticeK/F=5} on a Lattice. We compare the multi-LRU-One/All performance with different existing policies mentioned in this paper, like LFU, single-LRU, PBP \cite{BlaGioICC15} and GFI \cite{GolrezaeiINFOCOM12}, as well as the upper bound given in (\ref{UBhit}). The parameter $\alpha$ is chosen equal to $1\%$ in Fig.  \ref{fig:PoissonK/F=1}, \ref{fig:LatticeK/F=1} and $5\%$ in Fig. \ref{fig:LatticeK/F=5}. In both $F=10000$, $\gamma = 0.78$.

As a reminder, the single-LRU policy is not influenced by multi-coverage. Each user can contact a single station, the one closest to the user. If the user request is cached in this memory, then there is a hit, otherwise  the 
object is fetched from the core network and inserted to the station's cache. 

From the three figures very interesting conclusions about the policies can be derived:

(i) Even for small values of coverage overlap (expected coverage number) a considerable increase in hit probability is achieved by using the multi-LRU policies, compared to the single-LRU. For $\alpha=1\%$, when $\overline{N_{BS}}=2$, the multi-LRU-One is $42\%$ (relative gain) above the single-LRU for Lattice placement and 35\% for PPP placement. A further increase of $\overline{N_{BS}}$ makes the gain even more apparent. For  $\overline{N_{BS}}=3$ the relative gains are 70\% and 60\%, respectively.

 (ii) For every value of $\overline{N_{BS}}$ the multi-LRU-One policy performs better that the multi-LRU-All, in all three figures. This is because, for stationary traffic, when the same object is inserted in all stations covering a user (case -All), a request falling in areas of overlap profits less by the diversity of content from the multiple covering stations. 
 
 
 (iii) From both figures, the difference in performance between POP \{LFU, PBP, GFI\} and POQ \{multi-LRU-One/All, simple LRU\} policies is evident. For IRM traffic, POP policies have greater performance by exploiting the "expensive" information of known object popularity, which is assumed constant over time. In a realistic environment however, where traffic patterns change with time, such policies will demand regular updates and are approximative, because they depend on estimations over the popularity values. On the other hand, the multi-LRU policies introduced here do not depend on such information. A related interesting remark is that, as the MCSR $\alpha$ increases, the difference between the two groups' performance decreases. This can be observed by comparing Fig. \ref{fig:LatticeK/F=5} to Fig. \ref{fig:LatticeK/F=1} (Lattice).
 
 (iv) For $\overline{N_{BS}}$ close to 1, a user can connect to approximately one station, and the performance of multi-LRU-One/All, and single-LRU coincide. The same applies for the group LFU, PBP and GFI. For $\overline{N_{BS}}\approx 1$ these last three policies tend to cache the K most popular objects in each station. Hence, when a user connects to a single station then she/he gets the maximum hit probability and the upper bound also coincides.
 
(v) It is obvious that the two standard policies single-LRU and LFU exhibit constant performance as the multi coverage event increases, because the memory of each station is updated independently of the others and a user is served by at most one station. 

 (vi) GFI performs best among all policies, and its performance is very close to the upper bound. The latter is an indication that the upper bound is fairly tight. The good performance of the GFI comes at the cost of very high computational complexity for the memory allocations, as well as centralised implementation that requests considerable amount of information availability. 


\subsubsection{q-LRU}

Fig. \ref{fig:PoissonK/F=1_qLRU} plots the hit probability of q-multi-LRU-All policies for various values of $q\in\left(0,1\right]$. As in the previous figures, $\gamma=0.78$, $F=10,000$, $\lambda_b=0.5$ and stations are
modelled by PPP and have memory $K=100$.

When $q=1=100\%$, q-multi-LRU-All reduces to the multi-LRU-All policy. 
As $q$ decreases, the performance of q-multi-LRU-All improves, but new content is inserted more rarely. In this sense, the good performance of q-multi-LRU-All with small $q$ exploits the IRM characteristic of stationary traffic, and will converge to good performance after a long transient period. This is often not realistic for traffic that exhibits faster variations in popularity and catalogue content.

\subsubsection{Zipf parameter $\gamma$}
We provide plots for the hit probability versus this parameter in Fig. \ref{fig:VaryGamma_K/F=1}. Letting $\gamma$ increase results in a popularity distribution where a small number of objects is considerably more popular than the rest of the catalogue. Eventually, hit probability will increase for both multi-LRU policies, because due to the Update phase, popular objects tend to be kept cached in memory once inserted and get hit more often. Furthermore, the relative difference $\frac{P_{hit}(multi-LRU-One)-P_{hit}(multi-LRU-All)}{P_{hit}(multi-LRU-All)}$ decreases as $\gamma$ increases. This happens because for increasing $\gamma$ unpopular objects have less influence on the hit probability. 
%
%

\subsubsection{Memory-to-Catalogue-Size Ratio MCSR $\alpha$}
Fig. \ref{fig:VaryK/F} illustrates  the behaviour of the three policies \{multi-LRU-One, multi-LRU-All, PBP\} when varying $\alpha$ (here a larger size catalogue of $F=20,000$ is used, in order to evaluate for very small values of the $\alpha = K/F$ ratio). The hit probability increases when the ratio  $\alpha$ increases, and tends to $100\%$ as the ratio tends to 1. Furthermore, the need for smart memory allocations is less important for large values of the ratio $\alpha=K/F$ because the sum popularity of files left outside the caches is not considerable. Thus, we reasonably see in the figure, that different policies tend to have the same performance for larger values of the ratio $\alpha$.

%
%

\subsection{Traffic with temporal locality}
\label{Sec:5_4_TempLoc}

The evaluation up to this point has been restricted to IRM user (request) traffic, and it was observed in the simulation figures that the multi-LRU-One performs better than the multi-LRU-All. This is because IRM is stationary, so, by letting the simulations run for a long time period the performance of the multi-LRU-One can converge to high hit probabilities. This however is generally not true for traffic that exhibits temporal locality, like the model we introduce in Section \ref{Sec:4_Temp} of our work (see also \cite{TraversoTranMult15,OlmosTEMPO14}).


\begin{figure*}[th!]    
\centering  
\label{CacheEval2}        
        \subfigure[Hit Prob. VS $\overline{N_{bs}}$ (various cache-size $K$).]{          
           \includegraphics[width=0.315\textwidth]{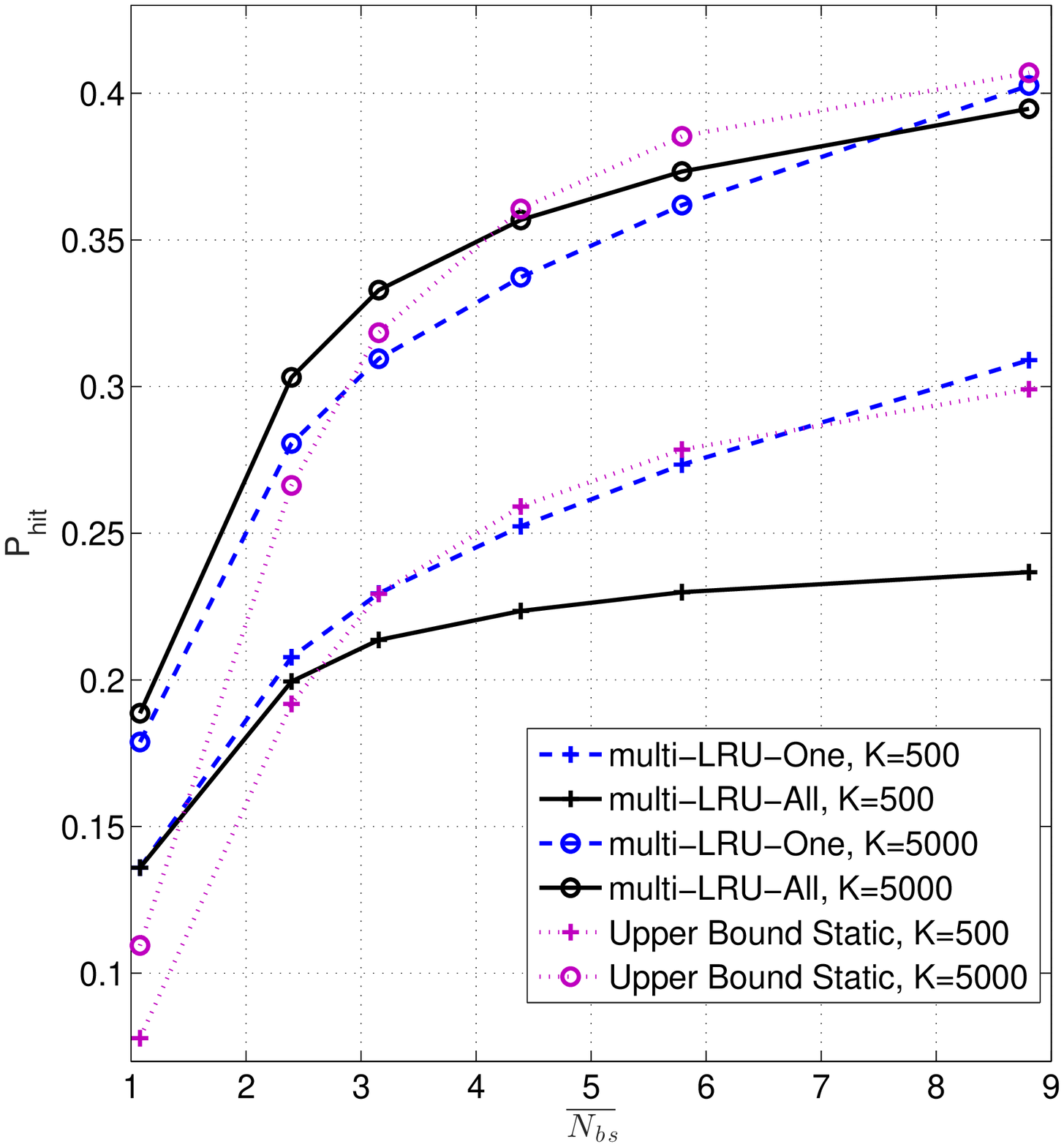}
           \label{fig:Hit_Nbs}
           }
	   \subfigure[Hit Prob. VS CCSR $\rho$ (various $\mathbb{E}{[V]}$).]{          
           \includegraphics[width=0.315\textwidth]{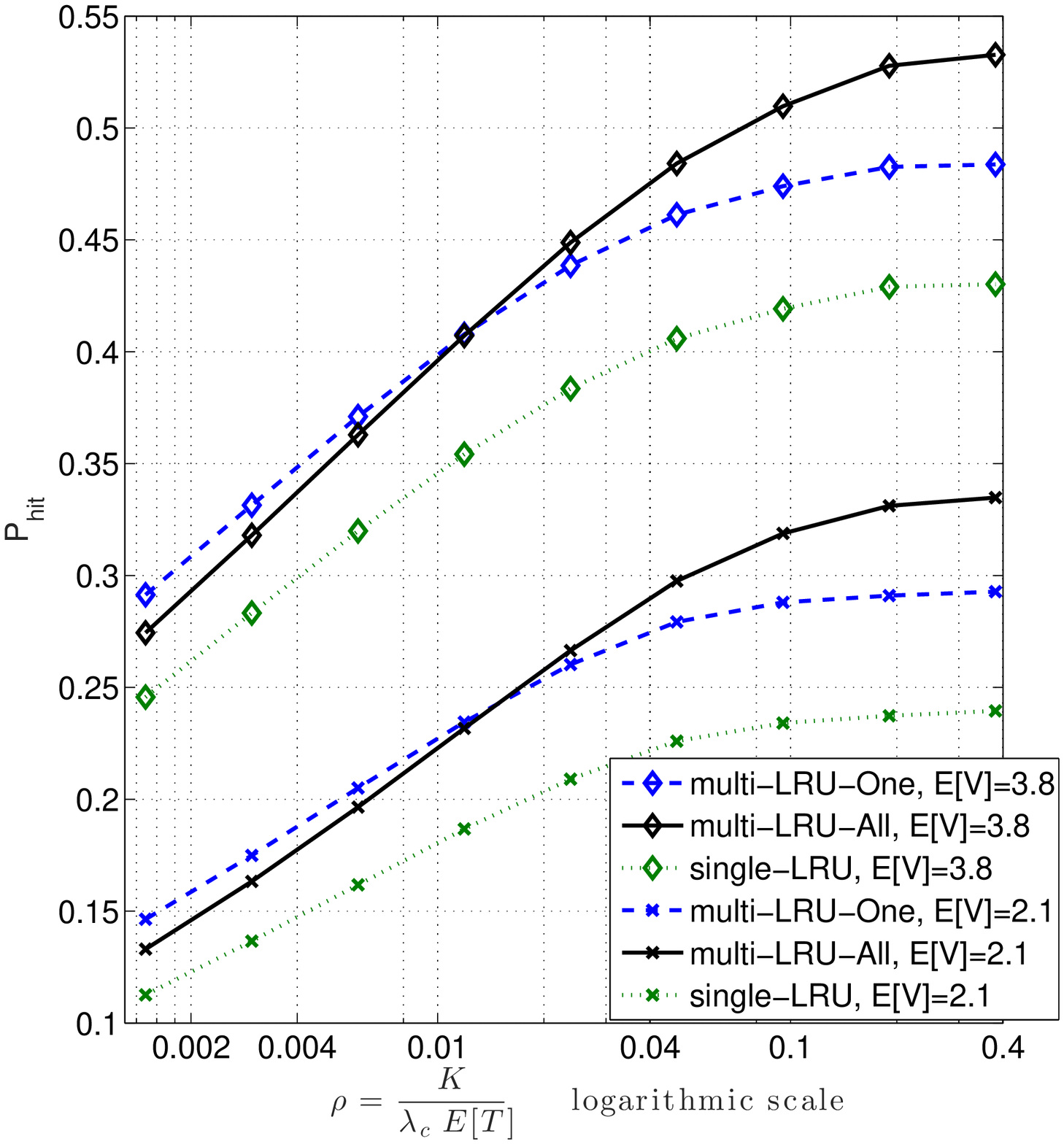}
           \label{fig:Hit_K}
           }
	   \subfigure[Hit Prob. VS $\overline{N_{bs}}$ (different popularity shapes).]{          
           \includegraphics[width=0.315\textwidth]{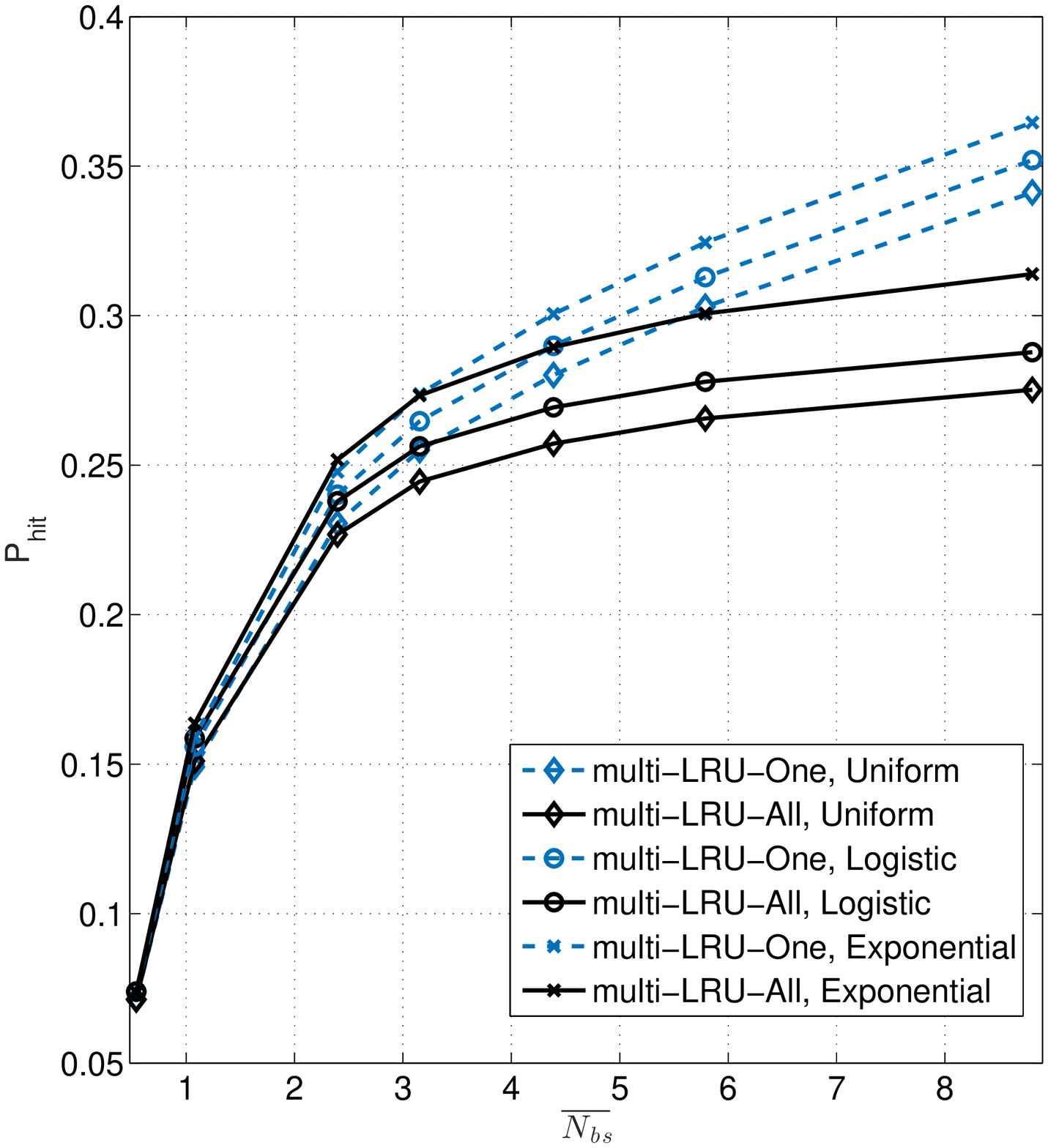}
           \label{fig:Hit_shapes}
           }
\caption{Evaluation of the hit probability of multi-LRU policies for different system variables. Parameter values in (b) $\overline{N_{bs}}=2.4$. In (c) $K=1500$.}
\end{figure*}
For the performance evaluation we consider $N_{st} = 20$ stations placed on a square lattice in a rectangular window, and the Boolean coverage model. 
If not mentioned otherwise, $\lambda_c=2400$ [objects/day], with expected request volume per content $\mathbb{E}[V]=2.1$.  The distribution lifespan p.d.f. is truncated Pareto, with $T\in[\tau_{min},\tau_{max}]= [0.1, 96]$ [$days$], and $\mathbb{E}[T] = 35$. The overall duration of the simulations is $15$ [$months$]. The popularity shape $g_i$ of content $c_i$ is chosen with probability $a_1=0.06$ as Logistic, $a_2=0.38$ as Gompertz and with $a_3=0.56$ as Exponential. Later we include the Uniform shape.

In each simulation, we keep some variables fixed and let others vary, to evaluate the policy performance and understand their influence. The variables to be varied are categorised into (1) \textit{network} variables (mean coverage number $\overline{N_{bs}}$, cache size $K$), and (2) \textit{traffic} variables (average request volume per content $\mathbb{E}[V]$, mean lifespan $\mathbb{E}[T]$, probability vector for the shape $(a_1, a_2, a_3)$).

\subsubsection{Network influence}



To increase the hit probability, data-objects should both: stay long-enough in each cache, and be inserted in as many caches as possible. But since storage space per station is limited, a trade-off arises, which is captured by the two variations of the multi-LRU policy (-One and -All). 
This trade-off is depicted in Fig. \ref{fig:Hit_Nbs} 
for different values of the memory size. The larger the storage space, the longer it takes for an object to be evicted. So, for large memory the geographical expansion of an object is beneficial. This is shown for $K=5000$ in which case multi-LRU-All surpasses multi-LRU-One. But for each $K$, there is a critical value of $\overline{N_{bs}}$ after which the performance of the -All is less than -One, because after this value further increase of content diversity is at the cost of content variety. The smaller the cache size, the more valuable storage space becomes because an object stays less time in the cache before eviction. Hence, for smaller $K$ -One shows better performance and exceeds the -All variation even for small values of multi-coverage $\overline{N_{bs}}$.

In Fig. \ref{fig:Hit_K}  the performance of multi-LRU policies versus the CCSR ratio $\rho$ (\ref{CCSR}) is illustrated. This ratio is equal to the mean number of memory slots per active content and is a measure of the system's storage capability, because the smaller it is than one, the less storage resources are available. 
Keeping the denominator of $\rho$ constant, Fig. \ref{fig:Hit_K} shows the impact of the memory size
on the policy performance. Obviously, hit probability increases with $K$, but for smaller $K$, the -One variation is preferable to the -All, as explained also previously. There is again a critical point in $K$ after which the -All variation is preferable (for large storage). If the ratio $\rho$ is further increased, the performance gains are diminishing for both variations, and saturation occurs. 

\subsubsection{Traffic influence}

The qualitative impact of the mean lifespan value on the hit probability can be understood by reading Fig. \ref{fig:Hit_K} in the opposite direction of the x-axis, from right to left. Keeping the numerator constant, as $\rho$ decreases $\mathbb{E}[T]$ increases. This means that the same storage capacity serves a larger active catalogue size $\mathbb{E}[\mathcal{F}]$. Consequently the overall performance drops. Moreover, Fig. \ref{fig:Hit_K} illustrates that the hit probability improves as the average number of requests per content  $\mathbb{E}[V]$ increases. The reason is that for higher $\mathbb{E}[V]$ a content put in storage is requested and hit more times.

In Fig. \ref{fig:Hit_shapes} each curve corresponds to a scenario where contents are assumed to follow only one particular popularity shape. Specifically, either the logistic, or the negative exponential, or the uniform shape is used. In the negative exponential shape popularity takes big values in a short time period and then drops abruptly. A steep popularity shape implies that consecutive requests of the same content appear close to each other in time. This makes more probable the event that the content is not evicted before its next request happens. With this in mind, it can be understood that the negative exponential can lead to higher hit probabilities than the uniform shape. Interestingly, for isolated caches the authors in \cite{TraversoTranMult15}, \cite{OlmosTEMPO14}  state that the shape does not affect significantly the hit probability of LRU. In our model, this can be observed when $\overline{N_{bs}}$ takes small values so every user can connect to at most one station, and this observation is confirmed in Fig \ref{fig:Hit_shapes}. But as $\overline{N_{bs}}$ increases and multi-coverage effects appear, \textit{the multi-LRU performance depends considerably on the correlation between requests of the same content, and thus the shape of the popularity curves}.

\subsubsection{Comparison with the single-LRU}

Under single-LRU a user can access only one (the closest in this work) station's memory even when covered by more than one. As a result hit performance is independent of $\overline{N_{bs}}$ (provided coverage is enough so that a user is always covered by at least one station). Depriving the user of the ability to retrieve its content from all covering stations, strongly reduces the overall hit probability. In Fig. \ref{fig:Hit_K}, where $\overline{N_{bs}}=2.4$, both multi-LRU policies show relative gains compared to the single-LRU, for all values of $\rho$. The maximum gains reach 30\% when $\mathbb{E}[V]=2.1$ and 20\% when $\mathbb{E}[V]=3.8$.

\subsubsection{Comparison with centralised Policies with periodic Popularity updates and prefetching (POP)}

Fig. \ref{fig:Hit_Nbs} shows that even the upper bound in (\ref{POPub}) for POP policies with estimated popularity input does not surpass the appropriate multi-LRU policy in performance, except maybe for a small range of $\overline{N_{bs}}$. For static IRM traffic, we saw in Fig.\ref{fig:PoissonK/F=1}-\ref{fig:LatticeK/F=5} that multi-LRU performed lower than the centralised policies POP. But, under a temporal traffic model (which is also more realistic), the ability of multi-LRU policies to update at each request the caches, without the need to estimate the content popularities, results in a considerable performance boost. 

\label{secVI}

\section{Conclusions}

In this work we have introduced a novel family of spatial multi-LRU policies, which exploit multi-coverage events of wireless networks to increase the hit probability. Two main variations are investigated, the multi-LRU-One and the -All. Che-like approximations give results close to simulation values. The multi-LRU-One provides higher object diversity in neighbouring caches and performs better under IRM traffic. The multi-LRU-All instead, lets objects quickly spread geographically and makes them immediately available to many users. This variation is profitable for traffic with temporal locality. Hence, depending on the incoming traffic either policy can be recommended. Future work should explain more clearly how the performance of these policies is affected by the spatial and temporal locality characteristics of traffic.


\section*{Acknowledgment}

The authors kindly thank Jim Roberts (IRT-SystemX, INRIA) for helpful discussions that improved the current work.

%
\bibliographystyle{abbrv}

%
%

%
\appendix

\begin{figure}[t!]   
\centering  
\subfigure[multi-LRU-One]{
	\includegraphics[trim=1cm 1.cm 1.8cm 1.2cm, clip, width=0.225\textwidth]{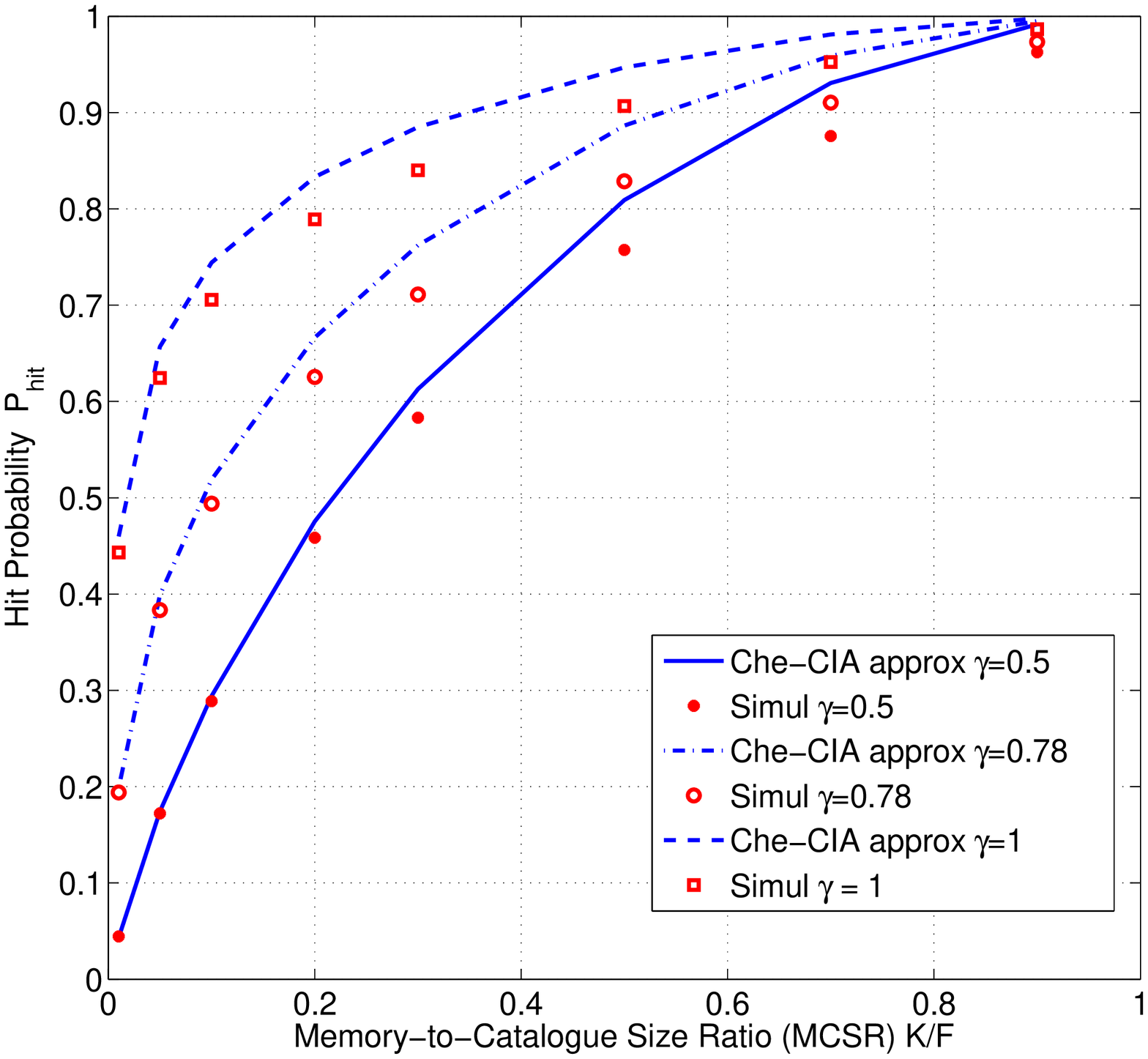}
           \label{Che2:One}
           }
           \subfigure[multi-LRU-All]{
           \includegraphics[trim=1cm 1.cm 1.8cm 1.2cm, clip, width=0.225\textwidth]{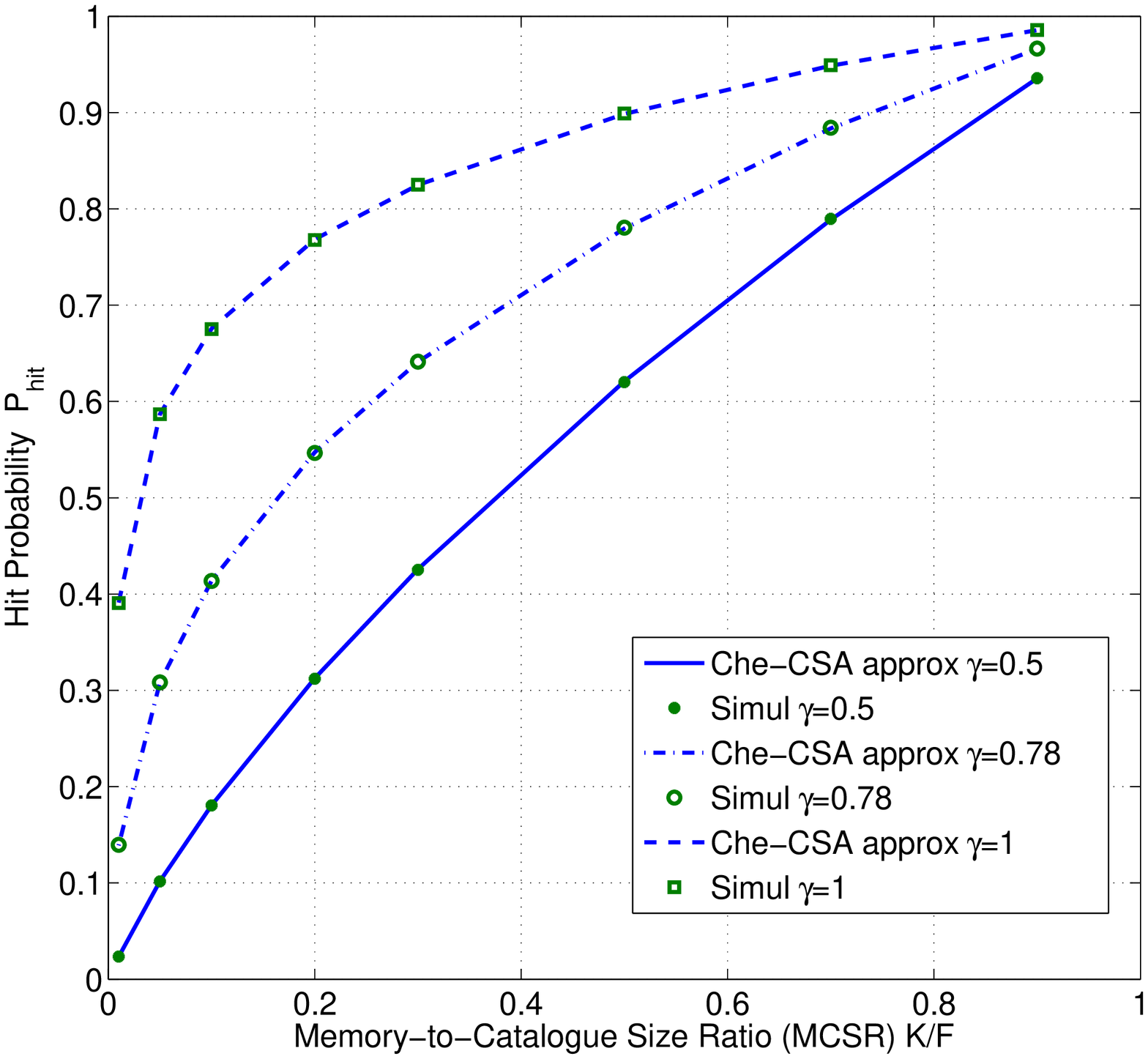}
	   \label{Che2:All}   
           }
         \caption{Che approximations for (a) multi-LRU-One with CIA and (b) multi-LRU-All with CSA, in the two-cache network. Hit probability versus MCSR $a=K/F$, $F=10,000$ objects, for different Zipf parameter $\gamma$.}
         \label{Che2cache}
\end{figure} 

\section{multi-LRU: two-cache network}
\label{app:B}

To understand how the Che-like approximations work for the multi-LRU policies, we analyse the simple network of two nodes $x_i$, $i\in\left\{1,2\right\}$, each one equipped with a cache of size $K$. 
Each node covers an entire area $A\subset\mathbb{R}^2$, so that all planar points are covered by both nodes. The total area is divided in two Voronoi cells $\mathcal{V}(x_i)$. To simplify further, we assume equal-sized Voronoi cells $|\mathcal{V}(x_1)|=|\mathcal{V}(x_2)|=|\mathcal{V}|$. 


We apply the analysis of Section \ref{GenChe} to this network model. Specifically, the formula for the hit probability of an object $c_j$ at cache $\Xi_i$ in (\ref{Phitgen2}), takes the expression (for $i=\{1,2\}$),
\begin{eqnarray}
\label{Phiti2}
P_{hit,i}\left(j\right) & \stackrel{IRM}{=} & \mathbb{P}\left(u_{-1}\in\left(\mathcal{S}_{-1},\left| t_{o}-t_{-1}\right|<T_{C},j\right)\right)\cdot \nonumber\\
& & \cdot \left[P_{hit,i}\left(j\right)+\mathbb{P}\left(c_j\notin\Xi_1\cap c_j\notin\Xi_2\right)\right].
\end{eqnarray} 
Solving the above over $P_{hit,i}\left(j\right)$ gives an expression for the hit probability of object $c_j$ at cache $\Xi_i$. The characteristic time $T_{C}$ is found by solving the equation (\ref{TCgen}),
\begin{eqnarray}
\label{Kincache}
\sum_{j=1}^F P_{hit,i}\left(j\right)=K, & i=\{1,2\}.
\end{eqnarray}
Finally, the total hit probability (\ref{PhitTOTgen}) takes both caches into account, and is equal to 
\begin{eqnarray}
\label{HitTot2}
P_{hit} & = & \sum_{j=1}^F a_j \left(1-\mathbb{P}\left(c_j\notin\Xi_1\cap c_j\notin\Xi_2\right)\right).
\end{eqnarray}

$\bullet$ \textbf{multi-LRU-One}: In this case, $\mathcal{S}_{o}=\mathcal{S}_{-1}=\mathcal{V}(x_i)$, in (\ref{Phiti2}). Table \ref{Tab2a} gives all pairs of inventory states that a user $u_o$ arriving at $t_o^{-}$ sees, when the previous user $u_{-1}$ asking for the same content arrived in cell (say) $\psi_{-1}\in\mathcal{V}(x_1)$ at some time $t_{-1}^{-}$, such that $\left|t_o-t_{-1}\right|\leq T_C$. We denote by logical $1$ the fact that the object is in the cache and by $0$ otherwise. From the table it is clear that user $u_{-1}$ does not take any action on cache $\Xi_2$, this is why, when $\mathbf{1}\left[c_j\in\Xi_2(t_{-1}^{-})\right]=1$, we cannot know whether the content will remain in the cache till $t_o^{-}$, so we write $\mathbf{1}\left[c_j\in\Xi_2(t_{o}^{-})\right]\in\left\{0,1\right\}$.
\begin{table}[ht!]
\caption{multi-LRU-One: States at $t_{-1}^-$ and $t_{o}^-$}
\centering
\begin{tabular}{| c | c | c | c | c | c | }
\hline
$\Xi_1(t_{-1}^{-})$ & $\Xi_2(t_{-1}^{-})$ &  & $\Xi_1(t_{o}^{-})$ & $\Xi_2(t_{o}^{-})$ & \\
\hline
0 & 0 & $\rightarrow$ & 1 & 0 & insert 1\\
0 & 1 & $\rightarrow$ & 0 & $\left\{0,1\right\}$ & no update\\
1 & 0 & $\rightarrow$ & 1 & 0 & update 1\\
1 & 1 & $\rightarrow$ & 1 & $\left\{0,1\right\}$ & update 1\\
\hline
\end{tabular}
\label{Tab2a}
\end{table}

There is the unknown probability $\mathbb{P}\left(c_j\neq \Xi_1\cap c_j\neq \Xi_2\right) = 1- \mathbb{P}\left(c_j \in\Xi_1\cup c_j\in\Xi_2\right)$. For multi-LRU-One, we observe that an insertion of an object is triggered when its request arrives but does not find the object inside any of the two caches. However, the insertion is done only in the closest cache and stays there for time $T_C$. During this time, the same object cannot be inserted in the other cache, hence, $\left\{c_j\in\Xi_1\right\}$ and $\left\{c_j\in\Xi_2\right\}$ are mutually exclusive events. Then,
\begin{eqnarray}
\label{jneq12one2}
\mathbb{P}\left(c_j\notin\Xi_1,c_j\notin\Xi_2\right) & = & 1 - \mathbb{P}\left(c_j \in\Xi_1\cup c_j\in\Xi_2\right)\nonumber\\ 
& = & 
1 - 2P_{hit,1}(j),
\end{eqnarray}
where the last equality is due to the symmetry of our model and the IRM traffic. However, in more general cases of node placement and coverage, content exclusivity is not true, because only a small area of the coverage cell will overlap with one neighbour. Users in other areas of the cell will be covered by other neighbours that can trigger the insertion of the same object, anyway. Hence, this result is not of much use for the PP coverage models. For this reason we want to evaluate how the CIA approximation applies here. For the two-cache model, this means for $\Xi_1$ (or $\Xi_2$),
\begin{eqnarray}
\label{CIA22}
\mathbb{P}\left(c_j\notin\Xi_1\cap c_j\notin\Xi_2\right) 
& = & 
 1 - P_{hit,1}(j), \ \ (CIA_1).
\end{eqnarray}
We can then replace in (\ref{Phiti2}) and (\ref{Kincache}) to get (for $i\in\left\{1,2\right\}$)
\begin{eqnarray}
\label{PhitOne2Ia}
P_{hit,i}(j) & = &  1-e^{-a_j\lambda_u|\mathcal{V}|T_C},\\
\label{PhitOne2Ib}
\sum_{j=1}^F P_{hit,i}(j) & = & \sum_{j=1}^F \left(1-e^{-a_j\lambda_u|\mathcal{V}|T_C}\right) = K.
\end{eqnarray}
For the total $P_{hit}$ probability, we should appropriately adapt the form in (\ref{HitTot2}) to the $CIA_2$ approximation,
\begin{eqnarray}
\label{PhitCIA2}
P_{hit} & = & \sum_{j=1}^F a_j \left(1-(\mathbb{P}\left(c_j\notin\Xi_1\right))^2\right)\nonumber\\
& = & \sum_{j=1}^F a_j (1-e^{-a_j\lambda_u2|\mathcal{V}|T_C}),
\end{eqnarray}
and the area $2|\mathcal{V}|=|A|$ is equal to the total coverage cell.


$\bullet$ \textbf{multi-LRU-All}: In this case, $\mathcal{S}_{-1}=\mathcal{S}_o= A$ in (\ref{Phiti2}) , for the hit probability of node $i$. 

To calculate the unknown probability $\mathbb{P}\left(c_j\notin\Xi_1\cap c_j\notin\Xi_2\right)$ we argue as follows. 
In the case of multi-LRU-All, an object cannot be inserted in cache 1 if not inserted also in cache 2 and the other way round. Based on the Che approximation, once the object is inserted it stays $T_C$ amount of time, before removed from each cache. Hence, the existence of an object in one cache implies the existence of the same object in the other. So, due to the model's symmetry
\begin{eqnarray}
\label{jneq12all2}
\mathbb{P}\left(c_j\notin\Xi_1\cap c_j\notin\Xi_2\right) & = & 1 - \mathbb{P}\left(c_j \in\Xi_1\cup c_j\in\Xi_2\right)\nonumber\\ 
& = & 
1 - P_{hit,1}(j).
\end{eqnarray}
This is simply the \textit{Cache Similarity Approximation (CSA)}, which for the two-cache network is exact! Then (\ref{Phiti2}) gives,
\begin{eqnarray}
\label{PhitiAll2}
P_{hit,i}(j) & = &
1-e^{-a_j\lambda_u|A|T_C}.
\end{eqnarray}
To find the characteristic time, we need to solve (\ref{Kincache}), 
\begin{eqnarray}
\label{TCAll2}
\sum_{j=1}^F P_{hit,i}(j)= \sum_{j=1}^F \left(1-e^{-a_j\lambda_u|A|T_C}\right) = K.
\end{eqnarray}
The total hit probability is equal to,
\begin{eqnarray}
\label{HitTotAll2}
P_{hit} & = & \sum_{j=1}^F a_j \left(1-\mathbb{P}\left(c_j\notin\Xi_1,c_j\notin\Xi_2\right)\right)\nonumber\\
& \stackrel{(\ref{PhitiAll2})}{=} & \sum_{j=1}^F a_j (1-e^{-a_j\lambda_u|A|T_C}).
\end{eqnarray}

An \textit{alternative way} to calculate $\mathbb{P}\left(c_j\notin\Xi_1(t_{o}),c_j\notin\Xi_2(t_{o})\right)$ is the following. A user $u_{o}$ finds the two caches without object $c_j$, if the previous user $u_{-1}$ (at say $\mathcal{S}_{-1}=\mathcal{V}(x_1)$) with the same demand, arrived either (i) at $t_{-1}^-:\left|t_{o}-t_{-1} \right|>T_C$, so that whatever the state of the two caches $\Xi_1(t_{-1}^-)$, $\Xi_2(t_{-1}^-)$, the object $c_j$ is eventually removed, since more than $T_C$ elapsed till $t_{o}$, or (ii) at $\left|t_{o}-t_{-1} \right|\leq T_C$. In the second case all possible change of states for the two caches is shown in Table \ref{Tab2b}. From this, we note that, the object will always be found in at least one of the two caches at $t_{o}$, so that the time difference can not be smaller than $T_C$. Hence,
\begin{eqnarray}
\label{jneq12all2alt}
\mathbb{P}\left(c_j\notin\Xi_1(t_{o}),c_j\notin\Xi_2(t_{o})\right) 
\stackrel{IRM}{=} e^{-a_j\lambda_u|A|T_C}.
\end{eqnarray}
The expressions in (\ref{jneq12all2alt}) and (\ref{PhitiAll2}) are the same.
\begin{table}[ht!]
\caption{multi-LRU-All: States at $t_{-1}^-$ and $t_{o}^-$}
\centering
\begin{tabular}{| c | c | c | c | c | c | }
\hline
$\Xi_1(t_{-1}^-)$ & $\Xi_2(t_{-1}^-)$ &  & $\Xi_1(t_{o}^-)$ & $\Xi_2(t_{o}^-)$ & \\
\hline
0 & 0 & $\rightarrow$ & 1 & 1 & insert both\\
0 & 1 & $\rightarrow$ & 0 & 1 & update 2\\
1 & 0 & $\rightarrow$ & 1 & 0 & update 1\\
1 & 1 & $\rightarrow$ & 1 & 1 & update both\\
\hline
\end{tabular}
\label{Tab2b}
\end{table}

The accuracy of the approximations in the two-cache network is shown in Fig.\ref{Che2cache}. The Che-CIA approximation for multi-LRU-One - although not accurate - performs reasonably well in the two-cache network. The Che-CSA approximation for the multi-LRU-All, is exact.

\end{document}